\documentclass[12pt,a4paper]{article}
\usepackage[dvips]{graphicx}
\usepackage{amssymb, amsmath, mathbbol}
\usepackage[centerlast,footnotesize]{caption2}
\usepackage{float}

\newcommand{\gapprox}{\raisebox{-0.5ex}{$\
\stackrel{\textstyle>}{\textstyle\sim}\ $}}

\usepackage{graphicx}
\newcommand{\One}{1\kern-4.5pt1}

\newcommand{\be}{\begin{equation}}
\newcommand{\ee}{\end{equation}}

\def\lesim{${\lower 2pt\hbox{$\scriptstyle
<$}\atop\raise 4pt\hbox{$\scriptstyle\sim$}}$} 
\def\grsim{${\lower2pt\hbox{$\scriptstyle >$} \atop\raise4pt\hbox 
{$\scriptstyle\sim$}}$} 
\begin{document}
\begin{center}
\begin{flushright}
July 2015
\end{flushright}
\vskip 10mm
{\LARGE
Domain Wall Fermions for Planar Physics
}
\vskip 0.3 cm
{\bf Simon Hands}
\vskip 0.3 cm
{\em Department of Physics, College of Science, Swansea University,\\
Singleton Park, Swansea SA2 8PP, United Kingdom.}
\end{center}

\noindent
{\bf Abstract:} 
In 2+1 dimensions, Dirac fermions in reducible, i.e. four-component
representations of the spinor algebra form the basis of many interesting model field
theories and effective descriptions of condensed matter phenomena.
This paper explores lattice formulations which preserve the global 
U$(2N_f)$ symmetry present in the massless limit, and its breakdown to 
U($N_f)\otimes$U($N_f$) implemented by three independent and parity-invariant fermion
mass terms. I set out generalisations of the Ginsparg-Wilson
relation, leading to a formulation of an overlap operator,
 and explore the remnants of the global symmetries which depart from
the continuum form by terms of order of the lattice spacing. I also define a
domain wall formulation in 2+1+1$d$, and present numerical evidence, in the form
of bilinear condensate and meson correlator calculations in quenched non-compact
QED using reformulations of all three mass terms, to show 
that U($2N_f$) symmetry is recovered in the limit that the domain-wall
separation $L_s\to\infty$.  The possibility that overlap and domain wall
formulations of reducible fermions may coincide only in the continuum limit is discussed.
\vspace{0.5cm}


\section{Introduction}

Relativistic fermions moving in two spatial dimensions are at the heart
of many interesting issues in theoretical physics, and are increasingly
important in effective descriptions of phenomena in condensed matter systems. 
Let us immediately draw a distinction between so-called {\it irreducible}
formulations in which the fermion fields $\psi,\bar\psi$  are described by two-component spinors, and {\it
reducible} models which invoke four-component spinors. In the former case,
there is no mass term invariant under a discrete parity inversion; gauge
theories of irreducible fermions generically manifest a parity
anomaly~\cite{Redlich:1983kn,Redlich:1983dv,Niemi:1983rq}, leading
to an induced Chern-Simons term endowing the gauge degrees of freedom with
mass. Such theories describe excitations with fractional statistics, and form
the basis for effective descriptions of the fractional quantum Hall
effect~\cite{Fradkin:1991wy}.
The main focus of this paper, by contrast, will be reducible theories, which like their 4$d$
counterparts admit
parity-invariant fermion masses, but whose action is invariant under
an enlarged group of global symmetries generated by both $\gamma_5$ and the
``unused'' $\gamma_3$.  The prototype model involving reducible fermions is
QED$_3$, an asymptotically-free theory with potential 
non-trivial critical dynamics in the infra-red, and still investigated as a toy
model of walking technicolor~\cite{Pisarski:1984dj}. Reducible fermions figure in effective
descriptions of the spin-liquid phase of quantum
antiferromagnets~\cite{Wen:2002zz,Rantner:2002zz}, the
pseudogap phase of cuprate superconductors~\cite{Herbut:2002yq,Franz:2002qy}, and
graphene~\cite{Khveshchenko:2001zz,Son:2007ja,CNGPNG}.

Once interactions are introduced, reducible 3$d$ fermions may exhibit spontaneous mass
generation via formation of a bilinear condensate $\langle\bar\psi\psi\rangle\not=0$, 
entirely analogous to chiral symmetry breaking in QCD. This is manifested in 
variants of the Gross-Neveu (GN) model~\cite{Rosenstein:1990nm}, in which the basic
interaction is a four-fermi contact between scalar or pseudoscalar densities; and
also in the Thirring model~\cite{Gomes:1990ed} whose interaction is a contact between two conserved currents, and
which shares the global symmetries of QED$_3$. For GN models the condensate
may be studied systematically using an expansion in $1/N_f$ where $N_f$ is the
number of fermion flavors~\cite{Hands:1992be}, whereas condensate formation in the Thirring model
is inherently non-perturbative.  In both cases the critical coupling at which
the condensate forms is associated with a UV-stable fixed point of the renormalisation
group, although the nature of the fixed-point theories remains an open
problem~\cite{Gehring:2015vja}.

Because field fluctuations near a fixed point can be large, it is natural to explore
critical dynamics in 3$d$ using lattice field theory techniques.
We immediately confront the issue of how best to
formulate the fermion fields. The two traditional approaches which mitigate the
species doubling present in naive discretisations are Wilson fermions, which
eliminate unwanted species by explicitly breaking important global symmetries,
and staggered fermions, which preserve a subgroup of the symmetries by
effectively spreading the spinor degrees of freedom over several lattice sites 
(for an excellent introduction to lattice fermions see chapters 5,7 and 10 of 
\cite{G&L}).  In a seminal work~\cite{Coste:1989wf}
Coste and L\"uscher argued that the Wilson formulation is the more natural for
irreducible descriptions, and can be shown to recover the correct parity anomaly (a
recent calculation of the mass-dependence of the anomaly in various background
fields using Wilson fermions is given in \cite{Karthik:2015sza}),
while the staggered approach naturally leads to a parity-invariant mass term;
indeed the relation between staggered fields and the $N_f=2$ reducible spinor
flavors expected in the continuum limit was
given by Burden and Burkitt in \cite{Burden:1986by}. Since the Chern-Simons
action is
imaginary in Euclidean metric, numerical simulations of $3d$ fermions have
to date been restricted almost exclusively to this second case.

To date there have been several papers studying critical behaviour in both GN
\cite{Hands:1992be,Karkkainen:1993ef,Focht:1995ie}
and Thirring~\cite{DelDebbio:1997dv,Frick:1994ry,Christofi:2007ye} 
models using numerical simulations of staggered fermions. In
general the results support the
theoretical prejudice that GN models exhibit a critical point at a coupling
$g^2\sim O(a)$ (where $a$ is lattice spacing) for all $N_f$, with small
corrections of 
O$(1/N_f)$ to both $g_c^2$ and the ``mean field'' critical exponents predicted in the large-$N_f$
limit~\cite{Hands:1992be}. By contrast, the Thirring model exhibits gap formation only for
$N_f\leq N_{fc}$, with both exponents and $g_c^2$ strongly
dependent on $N_f$~\cite{Christofi:2007ye}. In the continum, the two models are
supposedly distinct.  However, recent results obtained with a fermion bag
algorithm permitting simulation in the massless
limit~\cite{Chandrasekharan:2011mn,Chandrasekharan:2013aya} suggest that in fact
the lattice models defined using staggered fermions 
may actually lie in the same universality class for the minimal case $N_f=2$.
This somewhat unexpected result is motivation to question the applicability of
staggered fermions to problems involving
critical fields. For $N$ staggered flavors the global symmetry is
U($N)\otimes$U($N$), with $N_f=2N$, broken to U($N$) by the generation of a
fermion mass. In the long wavelength
limit $a\partial\to0$ this enlarges to the required
U($4N)\to$U($2N)\otimes$U($2N$)~\cite{Burden:1986by}, but there is less
reason than usual to trust this restoration once fluctuations on all length
scales are important. Other situations where it may be important 
to reproduce the global symmetry pattern correctly include the
infrared behaviour of QED$_3$, where it has been argued that dynamical mass
generation depends on the correct counting of massless degrees of freedom, which
include Goldstone modes in a symmetry-broken phase~\cite{Appelquist:1999hr}, and the role of
``half-instantons'' in the thermal response of the gapped phase of
graphene, which may result in a much reduced value for the critical
temperature~\cite{Aleiner:2007va}.

This paper explores the application of formulations originally 
developed to optimise
the reproduction of global symmetries in lattice QCD, namely
Ginsparg-Wilson (GW) fermions~\cite{Ginsparg:1981bj} and, principally, domain
wall fermions~\cite{Kaplan:1992bt,Furman:1994ky}, to 
reducible fermion models in 2+1$d$. After reviewing the relevant symmetries and
identifying three distinct but physically equivalent formulations of the mass term in
the next section, in Sec.~\ref{sec:GW} we generalise the GW
relation to fermions in 2+1$d$ and identify remnant quasi-global symmetries, which
recover the desired U(2$N_f$) form only in the continuum limit $a\to0$. A realisation of the GW
symmetries by an overlap operator~\cite{Neuberger:1997fp} is given. In Sec.~\ref{sec:DWF} we define a
domain wall fermion operator in 2+1+1$d$ which permits the definition of fermi
fields localised on domain walls at either end of the newly introduced
3 direction which purport to satisfy the U(2$N_f$) symmetry in the limit that the wall
separation $L_s\to\infty$. An important component of the argument is the
reformulation of the three distinct mass terms given in Sec.~\ref{sec:cont}.
Sec.~\ref{sec:numbers} presents results from numerical investigations of the
$N_f=1$ domain wall operator in the context of quenched non-compact QED$_3$, which
permits the use of either weak, strong, or intermediate coupling.  While there
is no attempt to explore either continuum or thermodynamic limits, we calculate
both bilinear condensates (Sec.~\ref{sec:condensates}) and meson propagators
(Sec.~\ref{sec:mesons}) using each of the three alternative mass terms, and show
that in almost all cases as $L_s\to\infty$ the results are in accord with a scenario in which U(2)
symmetry is
broken to U(1)$\otimes$U(1). Interestingly, the most rapid convergence to the
U(2)-symmetric limit is obtained for the case of a ``twisted'' mass term
$im\bar\psi\gamma_3\psi$.  For intermediate coupling the results for the
condensate $\langle\bar\psi\psi\rangle$ are compatible in the massless limit
with old results obtained with staggered fermions~\cite{Hands:1989mv}. Finally in
Sec.~\ref{sec:discussion} we present a summary of the findings 
and an outlook for future investigations. We also discuss the intriguing
possibility that for reducible theories of fermions in 2+1$d$ the overlap and
domain wall approaches may not coincide except in the continuum limit.

\section{Relativistic Fermions in 2+1d}
\label{sec:cont}

I begin by reviewing the continuum formulation of a gauge theory with fermion
fields $\Psi,\bar\Psi$ in a
reducible representation of the spinor algebra, based on $4\times4$ Euclidean Dirac matrices
$\gamma_\mu$ with $\{\gamma_\mu,\gamma_\nu\}=2\delta_{\mu\nu}$, $\mu,\nu=0,1,2$,  and having a
parity-invariant mass. The
weakly-interacting long-wavelength limit of staggered lattice fermions naturally
reproduces this formulation with $N_f=2$ flavors \cite{Burden:1986by} - in what follows
flavor indices are suppressed. The action can be written (for convenience, the
necessary $\int d^3x$ is omitted in all action definitions)
\begin{equation}
S=\bar\Psi D\Psi+m\bar\Psi\Psi
\label{eq:red_action}
\end{equation}
where the covariant derivative operator $D$ can be expanded as
\begin{equation}
D=\gamma_0 D_0+\gamma_1 D_1+\gamma_2 D_2=-D^\dagger.
\end{equation}
This has global symmetries 
\begin{eqnarray}
\Psi\mapsto e^{i\alpha}\Psi&;& 
\bar\Psi\mapsto\bar\Psi e^{-i\alpha},\label{eq:1}\\
\Psi\mapsto
e^{\alpha\gamma_3\gamma_5}\Psi&;&\bar\Psi\mapsto\bar\Psi
e^{-\alpha\gamma_3\gamma_5},\label{eq:g3g5}
\end{eqnarray}
where $\gamma_3$ and $\gamma_5$ are two additional traceless, hermitian, and
linearly independent 4$\times4$ matrices which anticommute with all the
$\gamma_\mu$ (see (\ref{eq:gammamu},\ref{eq:gamma35}) below), and
as usual in Euclidean matric $\gamma_5\equiv\gamma_0\gamma_1\gamma_2\gamma_3$.
For fermion mass $m=0$ there are two additional
symmetries
\begin{eqnarray}
\Psi\mapsto e^{i\alpha\gamma_5}\Psi&;&\bar\Psi\mapsto\bar\Psi
e^{i\alpha\gamma_5},\label{eq:g5}\\ 
\Psi\mapsto e^{i\alpha\gamma_3}\Psi&;&
\bar\Psi\mapsto\bar\Psi 
e^{i\alpha\gamma_3}.\label{eq:g3}
\end{eqnarray}
These four rotations generate a global U(2) invariance, which generalises to
U$(2N_f)$ for several flavors. The mass term explicitly 
breaks the symmetry from
U($2N_f)\to$ U$(N_f)\otimes$U$(N_f)$.

It will prove  interesting to  explore different forms of the mass term, which
are simply accessed by changing integration variables in the path integral. 
Since there is no axial anomaly in $2+1d$,
this procedure is straightforward in the continuum and the resulting action
describes
identical physics.  If, however,  the representations of the
Dirac matrices are tied to the particular form of the underlying lattice, as is
the case for staggered fermions or graphene, then due to discretisation effects 
the mass terms are not equivalent and correspond to
distinct patterns of symmetry breaking (see the discussion following
eqn.~(\ref{eq:act5}) for an example).
Let's recast the continuum action (\ref{eq:red_action}) in terms of two
two-component spinors $u$ and $d$:
\begin{equation}
S=\bar u\tilde Du-\bar d\tilde Dd +m\bar uu +m\bar dd.
\label{eq:irred_act}
\end{equation}
where $\tilde D=-\tilde D^\dagger=\sigma_1 D_0+\sigma_2 D_1+\sigma_3D_2$ and the $\sigma_i$ are Pauli
matrices. The link with (\ref{eq:red_action}) requires
the identification
\begin{equation}
\gamma_0=\left(\begin{matrix}\sigma_1&\cr&-\sigma_1\end{matrix}\right);\;\;
\gamma_1=\left(\begin{matrix}\sigma_2&\cr&-\sigma_2\end{matrix}\right);\;\;
\gamma_2=\left(\begin{matrix}\sigma_3&\cr&-\sigma_3\end{matrix}\right),
\label{eq:gammamu}
\end{equation}
implying
\begin{equation}
\gamma_3=\left(\begin{matrix}&-i\cr i&\end{matrix}\right);\;\;
\gamma_5=\left(\begin{matrix}&1\cr1&\end{matrix}\right);\;\;
i\gamma_3\gamma_5=\left(\begin{matrix}1&\cr&-1\end{matrix}\right).
\label{eq:gamma35}
\end{equation}
We now define an important discrete symmetry, parity, here specified  for
convenience in terms
of reversal of all three spacetime axes $x_\mu\mapsto-x_\mu$ (in general parity
must invert an odd number of axes, since flipping an even number is equivalent
to a rotation: the Euclidean parity operation which flips
just one axis is formally equivalent to the time-reversal operation frequently
discussed in condensed matter physics). In fact it can be realised in two ways:
\begin{eqnarray}
\bar u\mapsto \bar d;\;\;
\bar d\mapsto -\bar u;\;\;
u\mapsto d;\;\;
d\mapsto
-u;\;\;\;\;&\mbox{i.e.}&\;\Psi\mapsto
i\gamma_3\Psi;\;\;\bar\Psi\mapsto-i\bar\Psi\gamma_3
\label{eq:parity3}\\
\bar u\mapsto -i\bar d;\;\;
\bar d\mapsto -i\bar u;\;\;
u\mapsto id;\;\;
d\mapsto
iu;\;\;\;\;&\mbox{i.e.}&\;\Psi\mapsto
i\gamma_5\Psi;\;\;\bar\Psi\mapsto-i\bar\Psi\gamma_5\label{eq:parity5}
\end{eqnarray}
This should be no surprise, since both $\gamma_3$ and $\gamma_5$ behave
identically with respect to the $\gamma_\mu$ appearing in
(\ref{eq:red_action}). 
In either case the parity operation effectively exchanges the $u$ and $d$ fields, absorbing
the sign change of $\tilde D$ under $x\to-x$, but keeping the mass term invariant. 

Now consider a change of basis
\begin{equation}
\psi={1\over\surd2}(u+d);\;\;
\chi={1\over\surd2}(-u+d);\;\;
\bar\psi={1\over\surd2}(\bar u-\bar d);\;\;
\bar\chi={1\over\surd2}(\bar u+\bar d)
\label{eq:dwbasis}
\end{equation}
in which the action (\ref{eq:irred_act}) reads
\begin{equation}
S=(\begin{matrix}\bar\psi&\bar\chi\end{matrix})\left(\begin{matrix}\tilde D&-m\cr
m&-\tilde D\end{matrix}\right)\left(\begin{matrix}\psi\cr\chi\end{matrix}\right)
=(\begin{matrix}\bar\psi&\bar\chi\end{matrix})\left(\begin{matrix}\tilde D&-m\cr
m&\tilde D^\dagger\end{matrix}\right)\left(\begin{matrix}\psi\cr\chi\end{matrix}\right).
\label{eq:action}
\end{equation}
The parity transformation leaving (\ref{eq:action}) invariant is equivalent to
(\ref{eq:parity3}).
In terms of four-component spinors the action (\ref{eq:action}) can be written
\begin{equation}
S=\bar\Psi(D-im\gamma_3)\Psi.
\label{eq:act3}
\end{equation}
Although the mass term $im\bar\Psi\gamma_3\Psi$ is still parity invariant, it is
now an antihermitian term in the Lagrangian, in contrast to the hermitian mass of
(\ref{eq:red_action}). Superficially it resembles the {\it twisted\/} mass
sometimes used to 
formulate lattice QCD in 4$d$~\cite{Frezzotti:2000nk}, though the absence of an anomaly in $2+1d$
permits the term to be flavor singlet. 
We can also  consider a different change of basis
\begin{equation}
\psi={1\over\surd2}(u+id);\;\;
\omega={i\over\surd2}(u-id);\;\;
\bar\psi={1\over\surd2}(\bar u+i\bar d);\;\;
\bar\omega={i\over\surd2}(\bar u-i\bar d)
\label{eq:dw5basis}
\end{equation}
yielding
\begin{equation}
S=(\begin{matrix}\bar\psi&\bar\omega\end{matrix})\left(\begin{matrix}\tilde D&-im\cr
-im&-\tilde D\end{matrix}\right)\left(\begin{matrix}\psi\cr\omega\end{matrix}\right)
=(\begin{matrix}\bar\psi&\bar\omega\end{matrix})\left(\begin{matrix}\tilde D&-im\cr
-im&\tilde D^\dagger\end{matrix}\right)\left(\begin{matrix}\psi\cr\omega\end{matrix}\right),
\label{eq:action5}
\end{equation}
with this time a parity transformation (\ref{eq:parity5}).
In terms of four-component spinors (\ref{eq:action5}) can be written
\begin{equation}
S=\bar\Psi(D-im\gamma_5)\Psi;
\label{eq:act5}
\end{equation}
again, the mass term is parity invariant and antihermitian.

While the three actions (\ref{eq:red_action},\ref{eq:act3},\ref{eq:act5}) must
be equivalent in the continuum, when derived from a lattice system
such as a tight-binding model of graphene the terms correspond to physically distinct
gapping instabilities at the Dirac cones~\cite{Herbut:2009qb}. The mass term $mS_h$ of
(\ref{eq:red_action}) corresponds to a charge density wave in which electrons
preferentially sit on one of the two sub-lattices $A$ or
$B$~\cite{Semenoff:1984dq}, whereas a linear combination
of $m_3S_3$ (\ref{eq:act3}) and $m_5S_5$ (\ref{eq:act5}) yields a bond density wave in
which electrons are distributed on both $A$ and $B$ sublattices in a
Kekul\'e texture~\cite{Hou:2006qc}.

Note that a mass term proportional to the bilinear
$\bar\Psi\gamma_3\gamma_5\Psi$, whether hermitian or antihermitian, is
qualitatively different. In terms of two-component spinors the resulting action
reads
\begin{equation}
S=(\begin{matrix}\bar\psi&\bar\xi\end{matrix})\left(\begin{matrix}\tilde D+m&\cr
&-\tilde D-m\end{matrix}\right)\left(\begin{matrix}\psi\cr\xi\end{matrix}\right);
\label{eq:action35}
\end{equation}
since all elements are proportional to the combination $\tilde D+m$ 
there is no parity operation realisable as a linear transformation on
$\Psi,\bar\Psi$ which flips the sign of $\tilde D$ but leaves the mass
term invariant. This corresponds to the non-time-reversal invariant 
``Haldane mass'', realised in graphene-like systems by alternately circulating currents in
adjacent half-unit cells~\cite{Haldane:1988zza}.

In summary, there are three linearly independent, physically indistinguishable 
parity-invariant mass terms
available for continuum four-component Dirac fermions in 2+1$d$. We will see 
that this furnishes a non-trivial test for lattice fermion
formulations, such as the domain wall formulation presented in
Sec.~\ref{sec:DWF}, in which the matrices $\gamma_3$ and $\gamma_5$ appear in
inequivalent ways. Of course, the transformations (\ref{eq:dwbasis},\ref{eq:dw5basis})
are merely special cases of the rotations (\ref{eq:g3},\ref{eq:g5}) with
$\alpha={\pi\over4}$, so the physical equivalence of the mass terms is a
consequence of U(2) symmetry.

\section{Ginsparg-Wilson Relations and the Overlap operator}
\label{sec:GW}

The Nielsen-Ninomiya theorem \cite{Nielsen:1980rz,Nielsen:1981xu} famously asserts the impossibility of a lattice
fermion formulation with a physical (ie. undoubled) spectrum which simultaneously respects locality, unitarity (or
reflection positivity in Euclidean metric) and the existence of a conserved
axial charge, expressed in four dimensions via the anticommutator
$\{D,\gamma_5\}=0$.
Ginsparg and Wilson \cite{Ginsparg:1981bj} proposed a minimal modification to
these criteria for lattice chiral fermions to be viable, namely that chiral
symmetry now be constrained by the GW relation
\begin{equation}
\gamma_5 D+D\gamma_5=aD\gamma_5 D,\label{eq:GW5}
\end{equation}
where $a$ is the lattice spacing.
The non-vanishing right hand side of (\ref{eq:GW5}) is effectively an O($a$)
contact term for
the anticommutator of the propagator $D^{-1}$ with $\gamma_5$, which should
become unimportant in the long-wavelength limit $aD\to0$.  

The GW relation appropriate to odd-dimensional fermions in irreducible
representations of the spinor algebra was investigated by Bietenholz and
Nishimura~\cite{Bietenholz:2000ca}, who found that the correct parity anomaly is recovered due to non-invariance
of the fermion measure under a ``generalised parity'' transformation.
For the reducible 2+1$d$ theories of
interest in this paper the 
GW relations generalise to:
\begin{eqnarray}
\gamma_5 D+D\gamma_5&=&aD\gamma_5 D;\nonumber\\
\gamma_3 D+D\gamma_3&=&aD\gamma_3 D;\label{eq:GW3}\\
\gamma_3\gamma_5 D-D\gamma_3\gamma_5 &=& 0.
\label{eq:GW35}
\end{eqnarray}
In addition we require the hermiticity relations
\begin{equation}
\gamma_3D\gamma_3=\gamma_5D\gamma_5=D^\dagger.
\end{equation}

The GW symmetries (\ref{eq:GW5}-\ref{eq:GW35}) are realised by the overlap
operator~\cite{Neuberger:1997fp}
\begin{equation}
D^{ov}
=a^{-1}(1+\gamma_3\mbox{sign}[\gamma_3 A])
=a^{-1}(1+\gamma_5\mbox{sign}[\gamma_5 A])
=a^{-1}\left(1+{A\over{(A^\dagger
A)^{1\over2}}}\right),
\end{equation}
where the matrix $A\equiv aD^{Wils}-\One$, where $D^{Wils}$ is the 
lattice Dirac operator for orthodox Wilson fermions. It  has the properties
$\gamma_5A\gamma_5=\gamma_3A\gamma_3=A^\dagger$ and
$i\gamma_3\gamma_5Ai\gamma_3\gamma_5=A$,
implying $D^{ov}+D^{ov\dagger}=aD^{ov}D^{ov\dagger}=aD^{ov\dagger}D^{ov}$.
It is manifest that $\gamma_3$ and $\gamma_5$ are treated
identically in the overlap formalism. A distinct overlap
operator, appropriate for irreducible fermions and reproducing the correct parity
anomaly, was considered in \cite{Kikukawa:1997qh}.

Analogously to the situation in 4$d$~\cite{Luscher:1998pqa}
the GW relations (\ref{eq:GW5}-\ref{eq:GW35}) admit the following remnant symmetries,
which recover the desired U($2N_f$) symmetries (\ref{eq:1}-\ref{eq:g3}) in the
continuum and  long-wavelength limits $a\to0$, $aD\to0$:
\begin{eqnarray}
\Psi\mapsto e^{\left(i\alpha\gamma_5(1-{aD\over2})\right)}\Psi&;&
\bar\Psi\mapsto\bar\Psi
e^{\left(i\alpha(1-{aD\over2})\gamma_5\right)}\nonumber\\
\Psi\mapsto e^{\left(i\alpha\gamma_3(1-{aD\over2})\right)}\Psi&;&
\bar\Psi\mapsto\bar\Psi
e^{\left(i\alpha(1-{aD\over2})\gamma_3\right)}\nonumber\\
\Psi\mapsto e^{-\alpha\gamma_3\gamma_5}\Psi&;&\bar\Psi\mapsto\bar\Psi
e^{\alpha\gamma_3\gamma_5}.
\label{eq:remnant}
\end{eqnarray}
For theories such as the GN and Thirring models defined near a critical point it 
is not {\it a priori\/} clear that continuum and long-wavelength limits
coincide.

To construct mass terms, we define projection operators of two kinds
\begin{equation}
P_{5\pm}={1\over2}(1\pm\gamma_5);\;\;P_{3\pm}={1\over2}(1\pm\gamma_3),
\end{equation}
and 
\begin{equation}
\hat P_{5\pm}={1\over2}(1\pm\hat\gamma_5);\;\;\hat P_{3\pm}={1\over2}(1\pm\hat\gamma_3),
\end{equation}
with 
\begin{equation}
\hat\gamma_5=\gamma_5(1-aD);\;\;\hat\gamma_3=\gamma_3(1-aD).
\end{equation}
Relations (\ref{eq:GW5}-\ref{eq:GW35}) may be used to show $\hat\gamma_5^2=\hat\gamma_3^2=1$,
so that the $\hat P$s are genuine projectors. Importantly, it also follows that
\begin{equation}
D\hat P_{3\pm}=P_{3\mp}D;\;\;D\hat P_{5\pm}=P_{5\mp}D.
\end{equation}
Therefore it makes sense to define two different sets of projected fields:
\begin{eqnarray}
\Psi_{3\pm}&=&\hat P_{3\pm}\Psi;\;\;\bar\Psi_{3\pm}=\bar\Psi P_{3\mp};\nonumber\\
\Psi_{5\pm}&=&\hat P_{5\pm}\Psi;\;\;\bar\Psi_{5\pm}=\bar\Psi P_{5\mp},
\end{eqnarray}
enabling a decomposition of the kinetic term in two different ways:
\begin{equation}
\bar\Psi D\Psi=\bar\Psi_{3+}D\Psi_{3+}+\bar\Psi_{3-}D\Psi_{3-}=
\bar\Psi_{5+}D\Psi_{5+}+\bar\Psi_{5-}D\Psi_{5-}.
\end{equation}
The mass term in this approach is defined in terms of projected fields:
\begin{equation}
m_h(\bar\Psi_{3-}\Psi_{3+}+\bar\Psi_{3+}\Psi_{3-})=
m_h(\bar\Psi_{5-}\Psi_{5+}+\bar\Psi_{5+}\Psi_{5-})=m_h\bar\Psi(1-{aD\over2})\Psi\equiv
m_hS_h^{GW}.
\end{equation}
Although this term only corresponds with the desired mass term in the continuum
limit $a\to0$, the extra O($a$) piece has the same form as the kinetic term, so the
overall effect is a benign wavefunction renormalisation. We can also check the
effects of the GW symmetries (\ref{eq:remnant}), along with the fermion number
symmetry (\ref{eq:1}). With obvious notation, and to $O(\alpha)$:
\begin{eqnarray}
\delta_1^{GW}\bar\Psi(1-{aD\over2})\Psi&=&
\delta_{35}^{GW}\bar\Psi(1-{aD\over2})\Psi=0\\
\delta_{3,5}^{GW}\bar\Psi(1-{aD\over2})\Psi&=&2i\alpha\bar\Psi\gamma_{3,5}(1-{a^2D^\dagger
D\over4})\Psi\label{eq:GWGold1}
\end{eqnarray}
The variation (\ref{eq:GWGold1}) provides the interpolators for the Goldstone
modes associated with a spontaneous breaking
$\langle\bar\Psi(1-{aD\over2})\Psi\rangle\not=0$; we see there is an O($a^2$)
correction to the  expected continuum forms.

It is interesting to repeat this exercise for the two other parity-invariant
mass terms in (\ref{eq:act3},\ref{eq:act5}). The mass terms constructed from the
projected fields are
\begin{eqnarray}
im_3(\bar\Psi_{3-}\gamma_3\Psi_{3+}+\bar\Psi_{3+}\gamma_3\Psi_{3-})=
im_3(\bar\Psi_{5-}\gamma_3\Psi_{5-}+\bar\Psi_{5+}\gamma_3\Psi_{5+})&=&
im_3\bar\Psi\gamma_3(1-{aD\over2})\Psi\nonumber\\&\equiv& m_3S_3^{GW};\nonumber\\
im_5(\bar\Psi_{3-}\gamma_5\Psi_{3-}+\bar\Psi_{3+}\gamma_5\Psi_{3+})=
im_5(\bar\Psi_{5-}\gamma_5\Psi_{5+}+\bar\Psi_{5+}\gamma_5\Psi_{5-})&=&
im_5\bar\Psi\gamma_5(1-{aD\over2})\Psi\nonumber\\&\equiv& m_5S_5^{GW}.
\label{eq:GWtwist}
\end{eqnarray}
In this case the O($a$) correction does not correspond to an existing term in
the action, so the correction may be less benign. Another symptom
is that if we label the full Dirac
operator including mass terms as ${\cal D}=D+m_hS_1+m_3S_3+m_5S_5$, then
\begin{equation}
\gamma_3{\cal D}(m_h,0,0)\gamma_3={\cal D}^\dagger(m_h,0,0);\;\;\;
\gamma_5{\cal D}(m_h,0,0)\gamma_5={\cal D}^\dagger(m_h,0,0),
\label{eq:overlapdagger}
\end{equation}
but that no such simple relations exist once $m_3,m_5\not=0$.

We find the O($\alpha$) variations of the mass term $S_3$  under (\ref{eq:remnant})
\begin{eqnarray}
\delta_1^{GW}\bar\Psi\gamma_3(1-{aD\over2})\Psi&=&0\nonumber\\
\delta_3^{GW}\bar\Psi\gamma_3(1-{aD\over2})\Psi&=&2i\alpha\bar\Psi(1-{a^2DD^\dagger\over4})(1-{aD\over2})\Psi
\nonumber\\
\delta_5^{GW}\bar\Psi\gamma_3(1-{aD\over2})\Psi&=&i\alpha\bar\Psi\gamma_3\gamma_5{a\over2}(D-D^\dagger)
(1-{aD\over2})\Psi\nonumber\\
\delta_{35}^{GW}\bar\Psi\gamma_3(1-{aD\over2})\Psi&=&-2\alpha\bar\Psi\gamma_5
(1-{aD\over2})\Psi\label{eq:GWGold3}
\end{eqnarray}
with similar results, {\it mutatis mutandis\/} for variations of
$S_5$. In all cases the expected expressions for
Goldstone interpolators are recovered in the continuum limit. However for $a>0$
there are subtle differences in the effect of the GW variations on the different
mass terms; for instance, the two Goldstone interpolators ``$\bar\psi\psi$'' and
``$\bar\psi\gamma_5\psi$'' resulting from $\delta_3^{GW},\delta_{35}^{GW}$ in (\ref{eq:GWGold3}) 
have realisations differing by O($a^2$) away
from the continuum limit. In this sense, full U(2) symmetry is only
recovered by overlap fermions as $a\to0$.

\section{Domain Wall Formulation}
\label{sec:DWF}

In this section we set out an alternative route to undoubled U($2N_f$)-symmetric fermions
using the domain wall approach first introduced by Kaplan~\cite{Kaplan:1992bt} and put into the
present form by Furman and Shamir~\cite{Furman:1994ky}. Here we follow the treatment set out in
\cite{G&L}. 
Define 4-spinor fields $\Psi(x,s)$,
$\bar\Psi(x,s)$ living on a $4d$ lattice where $x$ denotes the usual 2+1$d$ spacetime
coordinates of a lattice site and $s=1,\ldots,L_s$ its coordinate along the
extra dimension, here labelled 3. 
The kinetic term in the action is then
\begin{equation}
S^{DW}=\sum_{x,y}\sum_{s,r}\bar\Psi(x,s)D^{DW}(x,s\vert y,r)\Psi(y,r).
\label{eq:SDWF}
\end{equation}
with domain wall Dirac operator 
\begin{equation}
D^{DW}(x,s\vert y,r)=\delta_{s,r}D(x\vert y)+\delta_{x,y}D_3^{DW}(s\vert r).
\end{equation}
The first term is the orthodox $2+1d$ Wilson operator 
\begin{equation}
D(x\vert y)={1\over2}\sum_{\mu=0,1,2}
\left[(1-\gamma_\mu)U_\mu(x)\delta_{x+\hat\mu,y}+(1+\gamma_\mu)U^\dagger_\mu(y)\delta_{x-\hat\mu,y}\right]
+(M-3)\delta_{x,y}
\label{eq:Ddw}
\end{equation}
with gauge link variables $U_\mu(x)$,
and $D_3^{DW}$ controls hopping in the 3 direction:
\begin{equation}
D_3^{DW}(s\vert
s^\prime)={1\over2}\bigl[(1-\gamma_3)\delta_{s+1,s^\prime}
(1-\delta_{s^\prime,L_s})
+(1+\gamma_3)\delta_{s-1,s^\prime}(1-\delta_{s^\prime,1})
-2\delta_{s,s^\prime}\bigr].
\label{eq:D3dw}
\end{equation}
Note there are Dirichlet boundary conditions imposed in direction 3, 
at $s=1$ and $s=L_s$. The inclusion of $D_3^{DW}$ explicitly
destroys the equivalence of $\gamma_3$ and $\gamma_5$ in the dynamics described
by the action (\ref{eq:SDWF}), so it
will be important to test whether and how this is recovered in practice. 

The key idea~\cite{Kaplan:1992bt}
is that the dynamics generated by (\ref{eq:Ddw}) and
(\ref{eq:D3dw}), with suitably chosen $M$, results in fermion zeromodes localised on 
{\it domain walls\/} at $s=1,L_s$, which are also respectively $\mp$ eigenmodes  of
$\gamma_3$. The $2+1d$ physics we wish to describe is formulated entirely using
these localised modes (the Wilson terms in
(\ref{eq:Ddw},\ref{eq:D3dw}) render the would-be zeromodes due to unwanted doubler
species non-normalisable in the limit $L_s\to\infty$~\cite{Kaplan:1992bt}). In particular we need to define $2+1d$
fermion mass terms corresponding to their continuum counterparts in
(\ref{eq:red_action}), (\ref{eq:act3}) and (\ref{eq:act5}). 
To this end, define fermion fields $\psi(x), \bar\psi(x)$ living in
2+1$d$:
\begin{eqnarray}
\psi(x)&=&P_-\Psi(x,1)+P_+\Psi(x,L_s);\nonumber\\
\bar\psi(x)&=&\bar\Psi(x,L_s)P_-+\bar\Psi(x,1)P_+,
\label{eq:3dfields}
\end{eqnarray}
where from now on $P_\pm\equiv{1\over2}(1\pm\gamma_3)$. 
We thus consider actions of the form
(\ref{eq:SDWF}) supplemented by three alternative mass terms:
\begin{eqnarray}
m_hS_h =
m_h\bar\psi\psi&=&m_h[\bar\Psi(x,L_s)P_-\Psi(x,1)+\bar\Psi(x,1)P_+\Psi(x,L_s)];\label{eq:dwmh}\\
m_3S_3 =
im_3\bar\psi\gamma_3\psi&=&
im_3[\bar\Psi(x,L_s)\gamma_3P_-\Psi(x,1)+\bar\Psi(x,1)\gamma_3P_+\Psi(x,L_s)];\label{eq:dwm3}\\
m_5S_5 =
im_5\bar\psi\gamma_5\psi&=&
im_5[\bar\Psi(x,L_s)\gamma_5P_+\Psi(x,L_s)+\bar\Psi(x,1)\gamma_5P_-\Psi(x,1)].\label{eq:dwm5}
\end{eqnarray}
It is interesting to note that $S_h$ has the same form as the fermion
mass term for domain wall formulations of $3+1d$ physics, and couples fields from opposite walls; $S_3$ also couples
opposite walls, but $S_5$ couples fields living on the same wall.

In the next section we will examine the numerical consequences of
the three terms (\ref{eq:dwmh}-\ref{eq:dwm5}) and in particular
check whether they yield compatible, U(2)-symmetric results in the
$L_s\to\infty$ limit.

\section{Numerical Results}
\label{sec:numbers}

In order to explore the domain wall action (\ref{eq:SDWF})
supplemented by one of the mass terms (\ref{eq:dwmh}-\ref{eq:dwm5}) we have
performed quenched simulations of non-compact QED, so that
$U_\mu(x)=\exp(i\theta_\mu(x))$ with the real link variable $\theta$ equilibrated
using the Maxwell-like action
\begin{equation}
S_{ncQED}={\beta\over2}\sum_{\mu\not=\nu}(\Delta^+_\mu\theta_\nu(x)-\Delta^+_\nu\theta_\mu(x))^2;
\;\;\;\mu,\nu=0,1,2.
\end{equation}
The dimensionless coupling $\beta\equiv(g^2a)^{-1}$, where the
dimensionful fermion charge $g$ is the natural scale with which to define physical
observables. The continuum limit lies at $\beta\to\infty$. In $2+1d$ QED is an
asymptotically-free theory whose infra-red behaviour is still imperfectly
understood. Since the non-compact lattice formulation is non-confining,
numerical simulations are plagued by the slow fall-off of the photon propagator
($\propto r^{-2}$ for the quenched theory),
and the thermodynamic limit is extremely difficult to achieve at weak coupling. 
Quenched simulations with staggered lattice fermions presented in
\cite{Hands:1989mv} suggest that chiral symmetry is spontaneously broken in the
ground state in the continuum thermodynamic limit, namely (in lattice units $a=1$)
\begin{equation}
\lim_{m\beta\to0}
\lim_{L/\beta\to\infty}
\lim_{\beta\to\infty}\langle\bar\psi\psi(m)\rangle\not=0.
\label{eq:chisb}
\end{equation}
Lattice sizes up to $80^3$ were used in support of this claim.

Away from the continuum limit massless staggered fermions have a manifest
U(1)$\otimes$ U(1) global symmetry which spontaneously breaks to U(1) either when
an explicit mass is introduced or a chiral condensate of the form
(\ref{eq:chisb}) develops. Only in the continuum limit can this pattern enlarge
to the expected U($2N_f)\to$U($N_f)\otimes$U($N_f$), with
$N_f=2$~\cite{Burden:1986by}.
The main aim of this study is not to verify (\ref{eq:chisb}) for domain wall
fermions, but rather to test the
restoration of the correct symmetry breaking pattern, with $N_f=1$, as
$L_s\to\infty$. To this end we have performed simulations on system sizes
$24^3\times L_s$, with couplings $\beta=0.5$, 1.0 and 2.0, corresponding
respectively to strong, intermediate and weak coupling regimes.  Throughout the
domain wall fermion mass $M$ in (\ref{eq:Ddw}) was set to 0.9 (physical
quantities are expected to be $M$-independent; §reflection
positivity requires $0<M<1$), and unless
otherwise stated the masses $m_h,m_3,m_5$ in (\ref{eq:dwmh}-\ref{eq:dwm5}) to
0.01. For convenience a hybrid Monte Carlo algorithm was used to generate the
$\{\theta\}$ ensembles, with 100 trajectories of average length 1.0 between
measurements; a much more efficient Fourier-space method was employed in
the original study \cite{Hands:1989mv}.

\subsection{Bilinear Condensates}
\label{sec:condensates}

First we explore bilinear condensates generically defined by 
$\langle\bar\psi\Gamma_i\psi\rangle\equiv\partial\ln{\cal
Z}/\partial m_i$, where $m_i\in\{m_h,m_3,m_5\}$. These form in 
response either to $m_i\not=0$ or in the thermodynamic limit to strong dynamics.
For the latter case in the limit $m_i\to0$ all three condensates should be equal if U(2) symmetry is
manifest. It is not obvious that this will occur
for the action (\ref{eq:SDWF}), both because of the Wilson
terms inherent in (\ref{eq:Ddw},\ref{eq:D3dw}), and because $\gamma_3$ and
$\gamma_5$ do not appear in (\ref{eq:SDWF}) on an equal footing.  By hypothesis,
however, U(2) symmetry should be recovered as $L_s\to\infty$.

To begin, we present results obtained using a spatial point source on a
configuation generated at $\beta=0.5$ (in fact, the numbers result from
averaging over 10 spatial sources); the systematics 
are easiest to expose at the strongest coupling. Note from
(\ref{eq:dwmh}-\ref{eq:dwm5}) that each condensate
gets contributions from two terms: for $\langle\bar\psi\psi\rangle$ and
$i\langle\bar\psi\gamma_3\psi\rangle$ the two terms arise from four-dimensional
propagators running from $s=1$ to $L_s$ and $L_s$ to 1 respectively; for
$i\langle\bar\psi\gamma_5\psi\rangle$ each contribution is from a propagator
starting and ending on the same domain wall. Within the working numerical
precison each contribution is the complex conjugate of the other, so the
sum is real. However, it turns out the imaginary component parametrises
the approach to the U(2)-symmetric limit. Define
$i\langle\bar\Psi(1)\gamma_3\Psi(L_s)\rangle
={i\over2}\langle\bar\psi\gamma_3\psi\rangle_{L_s}+i\Delta_h(L_s)$ (where the first
term is real, and the spatial coordinate $x$ is suppressed), and
then write:
\begin{eqnarray}
{1\over2}\langle\bar\psi\psi\rangle_{L_s}&=&{i\over2}\langle\bar\psi\gamma_3\psi\rangle_{L_S\to\infty}
+\Delta_h(L_s)+\epsilon_h(L_s);\\
{i\over2}\langle\bar\psi\gamma_3\psi\rangle_{L_s}&=&{i\over2}\langle\bar\psi\gamma_3\psi\rangle_{L_S\to\infty}
+\epsilon_3(L_s);\\
{i\over2}\langle\bar\psi\gamma_5\psi\rangle_{L_s}&=&{i\over2}\langle\bar\psi\gamma_3\psi\rangle_{L_S\to\infty}
+\epsilon_5(L_s).
\end{eqnarray}
The residuals $\Delta_h$ and $\epsilon_i$ must vanish for a
U(2)-invariant limit.

\begin{figure}[H]
\begin{center}
    \includegraphics[width=8.5cm]{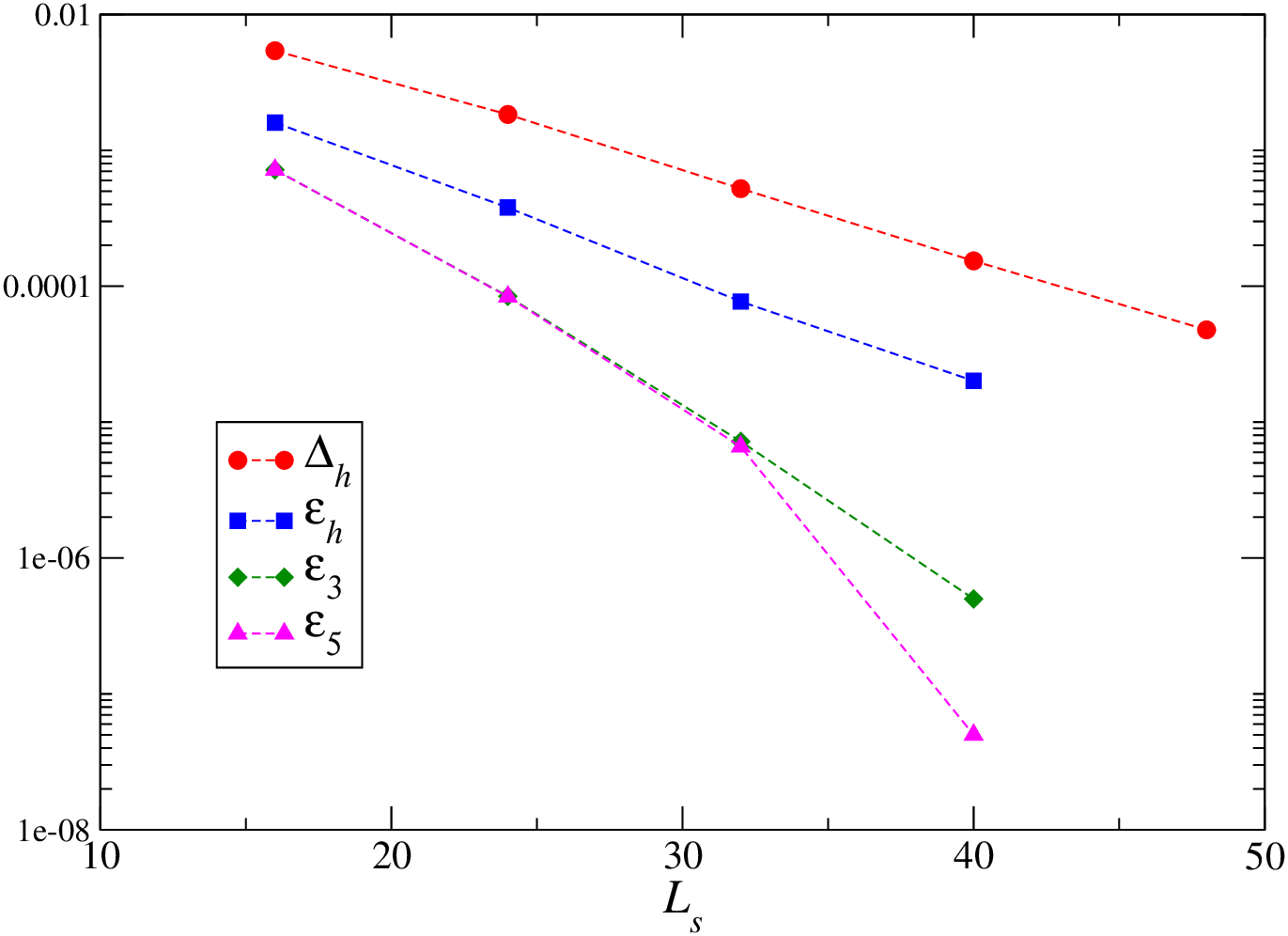}
\caption{Residual errors as a function of $L_s$ for bilinear condensates
evaluated using point spatial sources on $24^3$ at $\beta=0.5$.}
\label{fig:Delta_Ls}
\end{center}
\end{figure}
\begin{figure}[H]
\begin{center}
    \includegraphics[width=8.5cm]{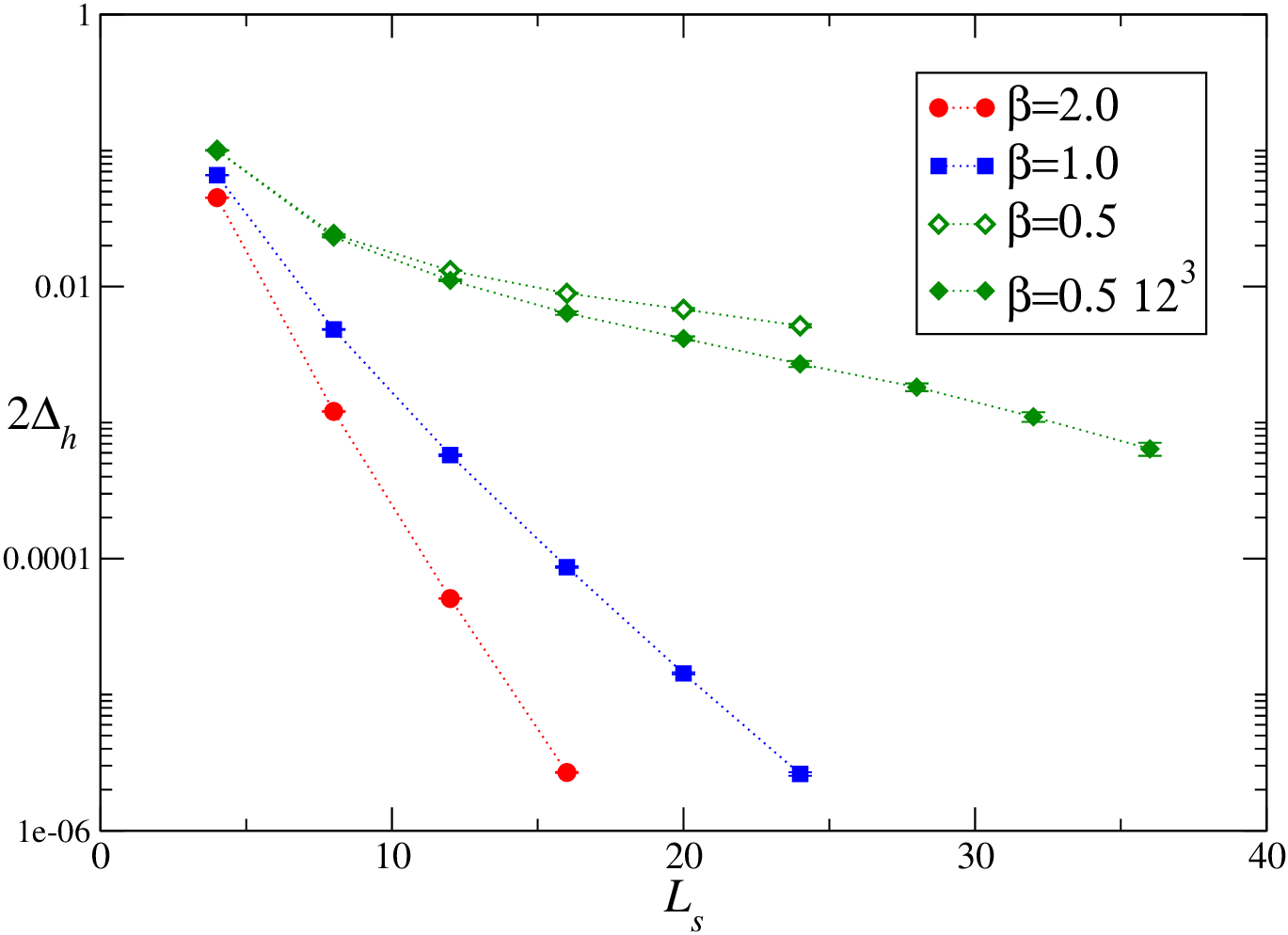}
\caption{Residual error $2\Delta_h(L_s)$ 
evaluated for various $\beta$ using a stochastic estimator on $24^3$.}
\label{fig:delta_beta}
\end{center}
\end{figure}
Fig.~\ref{fig:Delta_Ls} plots the residuals for $L_s=16,\ldots,40$; note that
$\Delta_h$ is measured directly as the imaginary component of
$i\langle\bar\psi\gamma_3\psi\rangle$ using just the $+$ components of
$\Psi,\bar\Psi$, while to estimate
the $\epsilon_i$ the value of
$i\langle\bar\psi\gamma_3\psi\rangle_{L_s\to\infty}$ is taken to be that measured at $L_s=48$.
Several features are apparent:

\begin{itemize}

\item
The dominant correction by almost an order of magnitude is $\Delta_h$, which
contributes to the hermitian condensate $\langle\bar\psi\psi\rangle$ but not, as
a result of the twist,  to
the antihermitian $i\langle\bar\psi\gamma_{3,5}\psi\rangle$. Indeed, at the
weakest coupling $\beta=2.0$
$\Delta_h$ is the only residual large enough to measure.

\item
All residuals decay approximately as $\exp(-cL_s)$, 
consistent with U(2) symmetry restoration as
$L_s\to\infty$.

\item
$\epsilon_3$ and $\epsilon_5$ are practically identical -- the disparity in
Fig.~\ref{fig:Delta_Ls} at $L_s=40$ probably results from uncertainty in the
determination of $i\langle\bar\psi\gamma_3\psi\rangle_{L_S\to\infty}$. This is
striking since the underlying structures in terms of four-dimensional propagators are quite
distinct.

\end{itemize}

Of course, we should not draw universal conclusions from a single gauge
configuration, particularly since the bilinear condensates display strong intermittent
upward fluctuations across a gauge ensemble. Fig.~\ref{fig:delta_beta} shows $\Delta_h(L_s)$ evaluated for
1200 ($24^3$) or 2400 ($12^3$) measurements of $i\langle\bar\psi\gamma_3\psi\rangle$ 
at the three couplings explored (for $L_s\geq16$ O(500) measurements were made
at the strongest coupling $\beta=0.5$ on $24^3$)
using a stochastic noise vector $\eta(x)$; to faithfully implement the projectors
$P_\pm$ in (\ref{eq:dwm3}) the same $\eta$ is chosen for all spin components
in the trace. Reassuringly, for $L_s\gapprox8$ the exponential behaviour $\Delta_h\propto
e^{-cL_s}$  persists, with decay constant $c$ decreasing as the coupling 
increases. Comparison of the $\beta=0.5$ results also shows $c$ is in general
also volume-dependent. Although not investigated here, we also expect $c$ to
depend on the Lagrangian parameter $M$. There is no reason to doubt, however, that the residuals will
eventually become an insignificant systematic as $L_s\to\infty$.
\begin{figure}[thb]
\begin{center}
    \includegraphics[width=8.5cm]{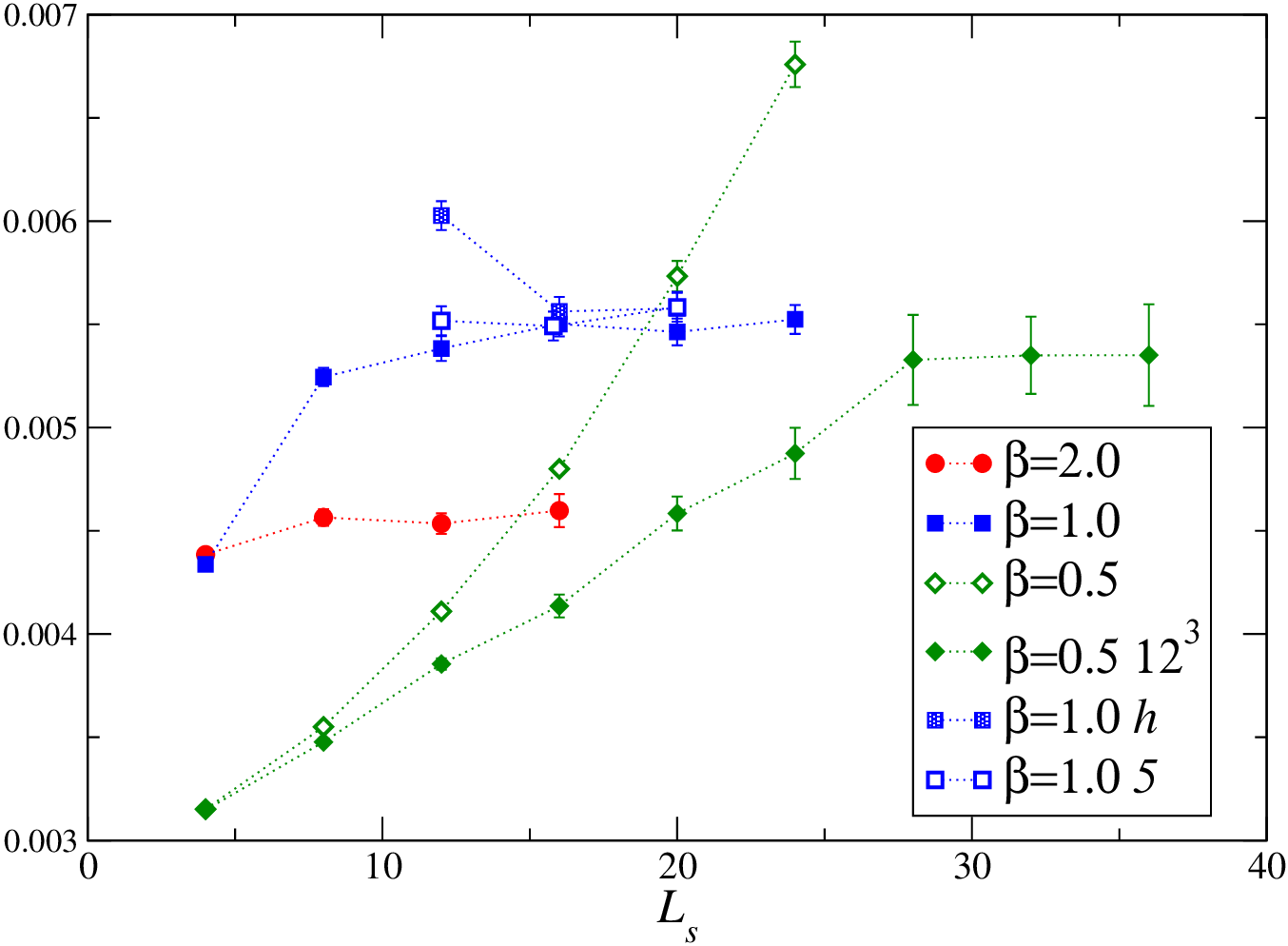}
\caption{Bilinear condensate $i\langle\bar\psi\gamma_3\psi\rangle$  for various $\beta$ 
as a function of $L_s$. Results for $\langle\bar\psi\psi\rangle$ and
$i\langle\bar\psi\gamma_5\psi\rangle$ at $\beta=1.0$ are also shown.}
\label{fig:cond_beta}
\end{center}
\end{figure}
\begin{figure}[!bh]
\begin{center}
    \includegraphics[width=8.5cm]{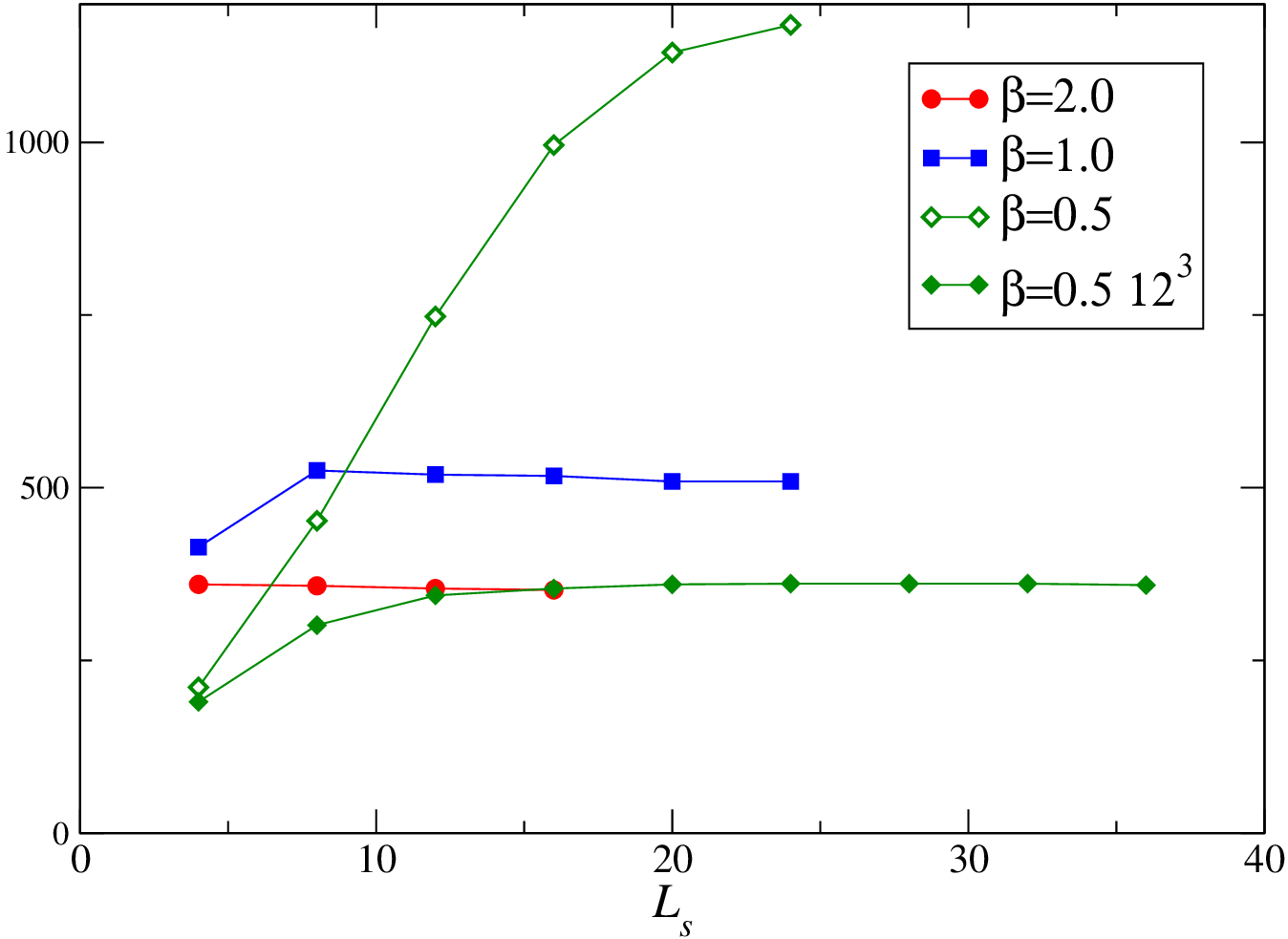}
\caption{Average number of conjugate gradient iterations per measurement
required for the data of Figs.~\ref{fig:delta_beta},\ref{fig:cond_beta}.}
\label{fig:ncg_beta}
\end{center}
\end{figure}
\begin{figure}[!bh]
\begin{center}
    \includegraphics[width=8.5cm]{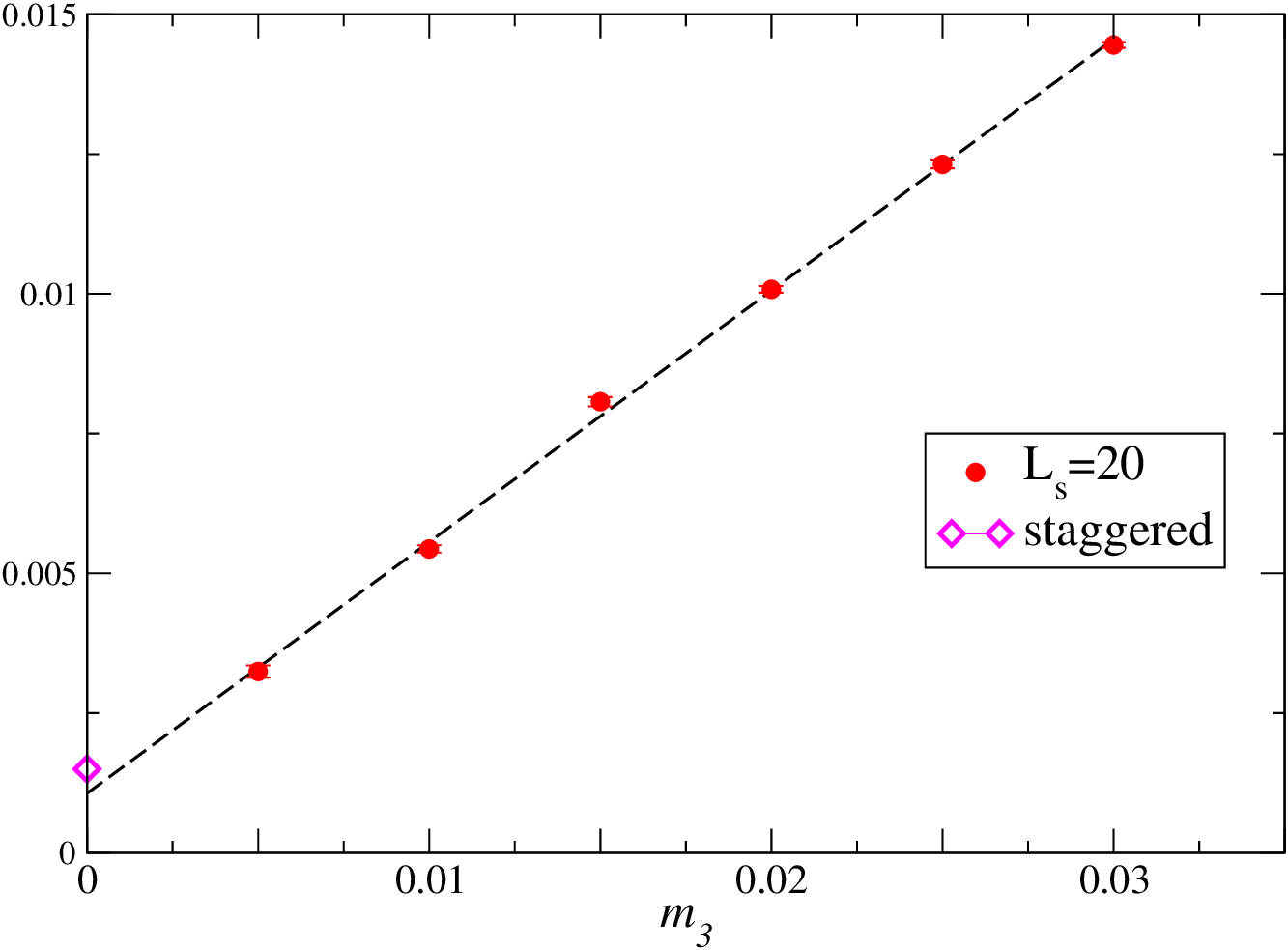}
\caption{$i\langle\bar\psi\gamma_3\psi\rangle$ vs. symmetry-breaking mass $m_3$
at $\beta=1.0$ on $24^3\times20$. The diamond shows the corresponding observable
extrapolated to the chiral limit using staggered fermions~\cite{Hands:1989mv}.}
\label{fig:cond_Ls20}
\end{center}
\end{figure}

Results for the condensate $i\langle\bar\psi\gamma_3\psi\rangle$, which as a
result of the twist does not include effects of the dominant residual
$\Delta_h$, are shown in Fig.~\ref{fig:cond_beta}. With the use of a stochastic
estimator a conjugate gradient residual of $10^{-12}$ per (four-dimensional)
lattice site and spin component sufficed. As a function of $L_s$ the
results plateau essentially once
$\Delta_h/\vert\langle\bar\psi\gamma_3\psi\rangle\vert$ is smaller than the
statistical error confirming the trends of Fig.~\ref{fig:delta_beta}: 
this occurs for $L_s\gapprox8$ for $\beta=2.0$ and
$L_s\gapprox16$ for $\beta=1.0$. At the strongest coupling $\beta=0.5$ the
available numerical resources have not permitted this regime to be probed on the
reference $24^3$ volume; however data taken on $12^3$ eventually plateau
for $L_s\gapprox28$. Based on the $\Delta_h$ data from Fig.~\ref{fig:delta_beta} we
might guess the $\beta=0.5$, $24^3$ plateau may set in for $L_s\approx40$.

Fig.~\ref{fig:cond_beta} also shows results for the other bilinear condensates
$\langle\bar\psi\psi\rangle$ ($h$) and $i\langle\bar\psi\gamma_5\psi\rangle$ ($5$) for
$\beta=1.0$, showing that within statistical errors they become consistent with
$i\langle\bar\psi\gamma_3\psi\rangle$ for $L_s\gapprox20$. The impact of the 
$\Delta_h$ residual on $\langle\bar\psi\psi\rangle$ is clearly discernable,
and suggests it is not the optimal choice for systematic studies of chiral
symmetry breaking. However, with this level of
precision there is no reason not to suppose U(2) symmetry is eventually restored. 
Fig.~\ref{fig:ncg_beta} shows the  number of conjugate gradient iterations
required for each fermion matrix inversion is roughly constant once the plateau is
reached, implying that the numerical effort needed scales linearly with $L_s$ as
the U(2)-symmetric limit is approached. 

Finally, Fig.~\ref{fig:cond_Ls20} plots $i\langle\bar\psi\gamma_3\psi\rangle$ as a function
of $m_3$ at $\beta=1.0$ on $24^3\times20$. The data can be plausibly extracted
linearly to the chiral limit $m_3\to0$ to yield a non-vanishing intercept,
consistent with the hypothesis that chiral symmetry is indeed spontaneously
broken (though at this stage no thermodynamic or continuum limit is taken). Also
shown is the staggered fermion condensate ${1\over2}\langle\bar\chi\chi\rangle$
(ie. normalised to
$N_f=1$) on the same spatial volume estimated from  Fig. 7 of
\cite{Hands:1989mv}. The values are close enough to be plausibly consistent, but
in any case provide reassurance first that the domain wall fermion data do not contain
significant contributions from doubler species, and secondly that the explicit
chiral symmetry-breaking violation by the Wilson terms in (\ref{eq:Ddw}) is mitigated in the
domain wall approach. Of course, since there are as-yet unquantified
renormalisations of both elementary fields and composite operators, these considerations
strictly require a continuum limit to be taken.

\subsection{Meson Propagators}
\label{sec:mesons}

Next we turn attention to correlations between mesonic operators
$\bar\psi\Gamma\psi$ at different spacetime points, with
$\Gamma\in\{\gamma_5,\gamma_3,\One, i\gamma_3\gamma_5\}$ all yielding spin-0
states, two with positive parity and two negative.
We again focus on the recovery of
U(2) symmetry as $L_s\to\infty$. A similar analysis for staggered
fermions in the continuum limit of non-compact QED$_3$ was presented in
\cite{Hands:2004bh}.
We begin by listing two important identities of the 2+1+1$d$ fermion propagator
$S(m_i;x,s;y,s^\prime)=\langle\Psi(x,s)\bar\Psi(y,s^\prime)\rangle$, which
follow directly 
from (\ref{eq:SDWF}) and have been checked numerically on small lattices:
\begin{eqnarray}
\gamma_5S(m_h,m_3,m_5;x,s;y,s^\prime)\gamma_5 &\equiv&
S^\dagger(m_h,m_3,-m_5;y,s^\prime;x,s);\label{eq:SvsSdag5}\\
\gamma_3S(m_h,m_3,m_5;x,s;y,s^\prime)\gamma_3 &\equiv& S^\dagger(m_h,-m_3,m_5;y,\bar s^\prime;x,\bar s).
\label{eq:SvsSdag3}
\end{eqnarray}
Here $x$ denotes the coordinate in 2+1$d$ and $s$ that in the 3rd direction,
$\bar s\equiv L_s-s+1$, and the dagger acts on Dirac indices.  
In what follows a mass parameter $m_i$ is omitted from the argument list if
it takes the value zero.
The identities (\ref{eq:SvsSdag5},\ref{eq:SvsSdag3}) are of course 
crucial for the efficient calculation of timeslice correlators with a minimal 
number of matrix inversions, although 
there will be  instances where a further inversion
with the sign of the mass term $m_{3,5}$ reversed is needed. The contrast with
relations (\ref{eq:overlapdagger}) for overlap fermions, which away from the
continuum limit only hold for $m_3=m_5=0$ should be noted. 
As will be demonstrated, 
in the limit $L_s\to\infty$ two further approximate relations become valid:
\begin{eqnarray}
\gamma_3S(m_3;x,s;y,s^\prime)\gamma_3&\approx&-S^\dagger(m_3;y,\bar
s^\prime;x,\bar s)\;\mbox{for}\;s^\prime=\bar s;\nonumber\\
\gamma_3S(m_3;x,s;y,s^\prime)\gamma_3&\approx& S^\dagger(m_3;y,\bar
s^\prime;x,\bar s)\;\mbox{for}\;s^\prime=s,\label{eq:LargeLsapprox}
\end{eqnarray}
and equivalent ones for $3\leftrightarrow5$.

Now consider the pion correlator
$C_5(x)=\langle\bar\psi(0)\gamma_5\psi\bar\psi(x)\gamma_5\psi(x)\rangle$. 
Under the parity definition (\ref{eq:parity3}) this state has
$J^P=0^-$.
Using the definition (\ref{eq:3dfields}) and ignoring 
the overall minus sign from the Grassmann nature of $\Psi,\bar\Psi$, we deduce the following
relation in terms of 2+1+1$d$ propagators
\begin{eqnarray}
C_5(x)&=&{\rm
tr}\Bigl[P_+S(m_i;x,L_s;0,L_s)P_-\gamma_5P_+S(m_i;0,L_s;x,L_s)P_-\gamma_5\nonumber\\
&+&P_-S(m_i;x,1;0,1)P_+\gamma_5P_-S(m_i;0,1;x,1)P_+\gamma_5\nonumber\\
&+&P_-S(m_i;x,1;0,L_s)P_-\gamma_5P_+S(m_i;0,L_s;x,1)P_+\gamma_5\nonumber\\
&+&P_+S(m_i;x,L_s;0,1)P_+\gamma_5P_-S(m_i;0,1;x,L_s)P_-\gamma_5\Bigr].
\end{eqnarray}
It's natural in this formalism to have projectors $P_\pm$ sandwiching the
fermion propagators --
they are the bridge, both formally and in the code, between the 4$d$ and 3$d$
worlds. 

Now specialise to the case of mass term $S_h$, and use the identity
(\ref{eq:SvsSdag5}), along with translational invariance in 3$d$, to arrive at
\begin{eqnarray}
C_5^h(x)&=&{\rm
tr}\Bigl[S(m_h;0,L_s;x,L_s)P_-S^\dagger(m_h;0,L_s;x,L_s)P_+\nonumber\\
&+&S(m_h;0,1;x,1)P_+S^\dagger(m_h;0,1;x,1)P_-\nonumber\\
&+&S(m_h;0,1;x,L_s)P_-S^\dagger(m_h;0,1;x,L_s)P_-\nonumber\\
&+&S(m_h;0,L_s;x,1)P_+S^\dagger(m_h;0,L_s;x,1)P_+\Bigr]\label{eq:C5}\\
&\equiv&C^{h-+}(x)+C^{h+-}(x)+C^{h--}(x)+C^{h++}(x).\label{eq:short}
\end{eqnarray}
This has the familiar look of a pion correlator (ie. positive definite terms of the form
$SS^\dagger$) apart from the projectors, and
is clearly calculable with two sources, one located at $s=1$ and the other at
$s=L_s$.
In the abbreviated form (\ref{eq:short}) note that $-+$ and $+-$
correlators involve fermion propagators linking a domain wall to itself, whereas 
for $--$ and $++$ the propagators run between the walls.

Next consider the $0^+$ correlator 
$C_3^h(x)=\langle\bar\psi(0)\gamma_3\psi\bar\psi(x)\gamma_3\psi(x)\rangle$. This
produces an expression with a different 4$d$ spacetime structure, but
which with the help of (\ref{eq:SvsSdag3}) can be rearranged
to coincide exactly with (\ref{eq:C5}): 
\begin{eqnarray}
C_3^h(x)&=&{\rm
tr}\Bigl[P_-S(m_h;x,1;0,L_s)P_-\gamma_3P_-S(m_h;0,1;x,L_s)P_-\gamma_3\nonumber\\
&+&P_-S(m_h;x,1;0,1)P_+\gamma_3P_+S(m_h;0,L_s;x,L_s)P_-\gamma_3\nonumber\\
&+&P_+S(m_h;x,L_s;0,L_s)P_-\gamma_3P_-S(m_h;0,1;x,1)P_+\gamma_3\nonumber\\
&+&P_+S(m_h;x,L_s;0,1)P_+\gamma_3P_+S(m_h;0,L_s;x,1)P_+\gamma_3\Bigr]\nonumber\\
&=&C^{h--}(x)+C^{h+-}(x)+C^{h-+}(x)+C^{h++}(x)\equiv C_5^h(x).\label{eq:C3}
\label{eq:Ch3Ch5}
\end{eqnarray}
These two states are thus exactly
degenerate if the hermitian mass term $S_h$ is chosen.

The same methods can be used for the correlators
\begin{eqnarray}
C_1(x)&=&\langle\bar\psi(0)\psi(0)\bar\psi(x)\psi(x)\rangle;\nonumber\\
C_{35}(x)&=&\langle\bar\psi(0)i\gamma_3\gamma_5\psi(0)\bar\psi(x)i\gamma_3\gamma_5\psi(x)\rangle,
\end{eqnarray}
which are respectively $0^+$ and $0^-$
to show 
\begin{eqnarray}
C_1^h(x)=C_{35}^h(x)&=&{\rm
tr}\Bigl[S(m_h;0,1;x,L_s)\gamma_3P_-S^\dagger(m_h;0,1;x,L_s)P_-\gamma_3\nonumber\\
&+&S(m_h;0,L_s;x,L_s)\gamma_3P_-S^\dagger(m_h;0,L_s;x,L_s)P_+\gamma_3\nonumber\\
&+&S(m_h;0,1;x,1)\gamma_3P_+S^\dagger(m_h;0,1;x,1)P_-\gamma_3\nonumber\\
&+&S(m_h;0,L_s;x,1)\gamma_3P_+S^\dagger(m_h;0,L_s;x,1)P_+\gamma_3\Bigr]\nonumber\\
&=&C^{h--}(x)-C^{h-+}(x)-C^{h+-}(x)+C^{h++}(x).
\label{eq:Ch1Ch35}
\end{eqnarray}
Therefore these two states are also degenerate, but distinct from
$\bar\psi\gamma_3\psi$ and $\bar\psi\gamma_5\psi$ in (\ref{eq:C5}). Assuming 
the primitive correlators $C^{h\pm\pm}$ are all positive definite, then 
$\vert C_5^h\vert=\vert C_3^h\vert>\vert C_1^h\vert=\vert C_{35}^h\vert$. 
If the symmetry breaking in the direction of 
$S_3$ is
spontaneous this is consistent with U(2)$\rightarrow$U(1)$\otimes$U(1) yielding
two light pseudo-Goldstone modes interpolated by $\bar\psi\gamma_5\psi$ and
$\bar\psi\gamma_3\psi$. Had we instead chosen (\ref{eq:parity5}) as the
definition of parity, then the assignments of the two Goldstone states would
be reversed so that $\bar\psi\gamma_5\psi$ is now $0^+$ and
$\bar\psi\gamma_3\psi$ $0^-$. However, the overall picture of a Goldstone
$0^\pm$ doublet and a non-Goldstone $0^\pm$ doublet remains unaltered.

The picture is more interesting when the mass term $S_3$ is considered. The
correlator 
$C_5^3=C^{3-+}+C^{3+-}+C^{3--}+C^{3++}$ in analogy to
(\ref{eq:C5},\ref{eq:short}), 
but for $C_3^3$
we find
\begin{eqnarray}
C_3^3(x)&=&{\rm
tr}\Bigl[S(m_3;0,1;x,L_s)P_-S^\dagger(-m_3;0,1;x,L_s)P_-\nonumber\\
&+&S(m_3;0,1;x,1)P_+S^\dagger(-m_3;0,1;x,1)P_-\nonumber\\
&+&S(m_3;0,L_s;x,L_s)P_-S^\dagger(-m_3;0,L_s;x,L_s)P_+\nonumber\\
&+&S(m_3;0,L_s;x,1)P_+S^\dagger(-m_3;0,L_s;x,1)P_+\Bigr]\\
&\equiv&\tilde C^{3--}(x)+\tilde C^{3+-}(x)+\tilde C^{3-+}(x)+\tilde C^{3++}(x).
\label{eq:C3m3}
\end{eqnarray}
The expression (\ref{eq:C3m3}) is similar in form to (\ref{eq:C3}) but requires
twice the number of matrix inversions to evaluate. 
However, use of the large-$L_s$
approximations (\ref{eq:LargeLsapprox}) gives
\begin{equation}
C_3^3(x)\approx-C^{3--}(x)+C^{3+-}(x)+C^{3-+}(x)-C^{3++}(x).
\end{equation}
Similarly we find
\begin{eqnarray}
C_1^3(x)&\approx&-C^{3--}(x)-C^{3++}(x)-C^{3-+}(x)-C^{3+-}(x);\\
C_{35}^3(x)&=&C^{3--}(x)+C^{3++}(x)-C^{3-+}(x)-C^{3+-}(x).
\label{eq:C35m3}
\end{eqnarray}
This suggests that the Goldstone modes are now interpolated by $\bar\psi\psi$
($0^+$) 
and $\bar\psi\gamma_5\psi$ ($0^-$), consistent with the variation of $S_3$ under
(\ref{eq:1}-\ref{eq:g3}), but that the
U(2)$\rightarrow$U(1)$\otimes$U(1) pattern will now only be recovered in the
limit $L_s\to\infty$. 

Finally, for mass term $S_5$ we have
\begin{eqnarray}
C_5^5(x)&\approx&C^{5-+}(x)+C^{5+-}(x)-C^{5--}(x)-C^{5++}(x);\label{eq:C55}\\
C_3^5(x)&=&C^{5-+}(x)+C^{5+-}(x)+C^{5--}(x)+C^{5++}(x);\label{eq:C35}\\
C_1^5(x)&=&-C^{5-+}(x)-C^{5+-}(x)+C^{5--}(x)+C^{5++}(x);\label{eq:C15}\\
C_{35}^5(x)&\approx&-C^{5-+}(x)-C^{5+-}(x)-C^{5--}(x)-C^{5++}(x);\label{eq:C355}
\end{eqnarray}
the apparent contradiction with the 
expected identification of $\bar\psi\psi$ ($0^+$) and
$\bar\psi\gamma_3\psi$ ($0^-$) (using either the symmetries (\ref{eq:1}-\ref{eq:g3}) or
the change of variables (\ref{eq:dw5basis}))
as Goldstone interpolators will be further discussed below.

\begin{figure}[tbh]
\begin{center}
    \includegraphics[width=10cm]{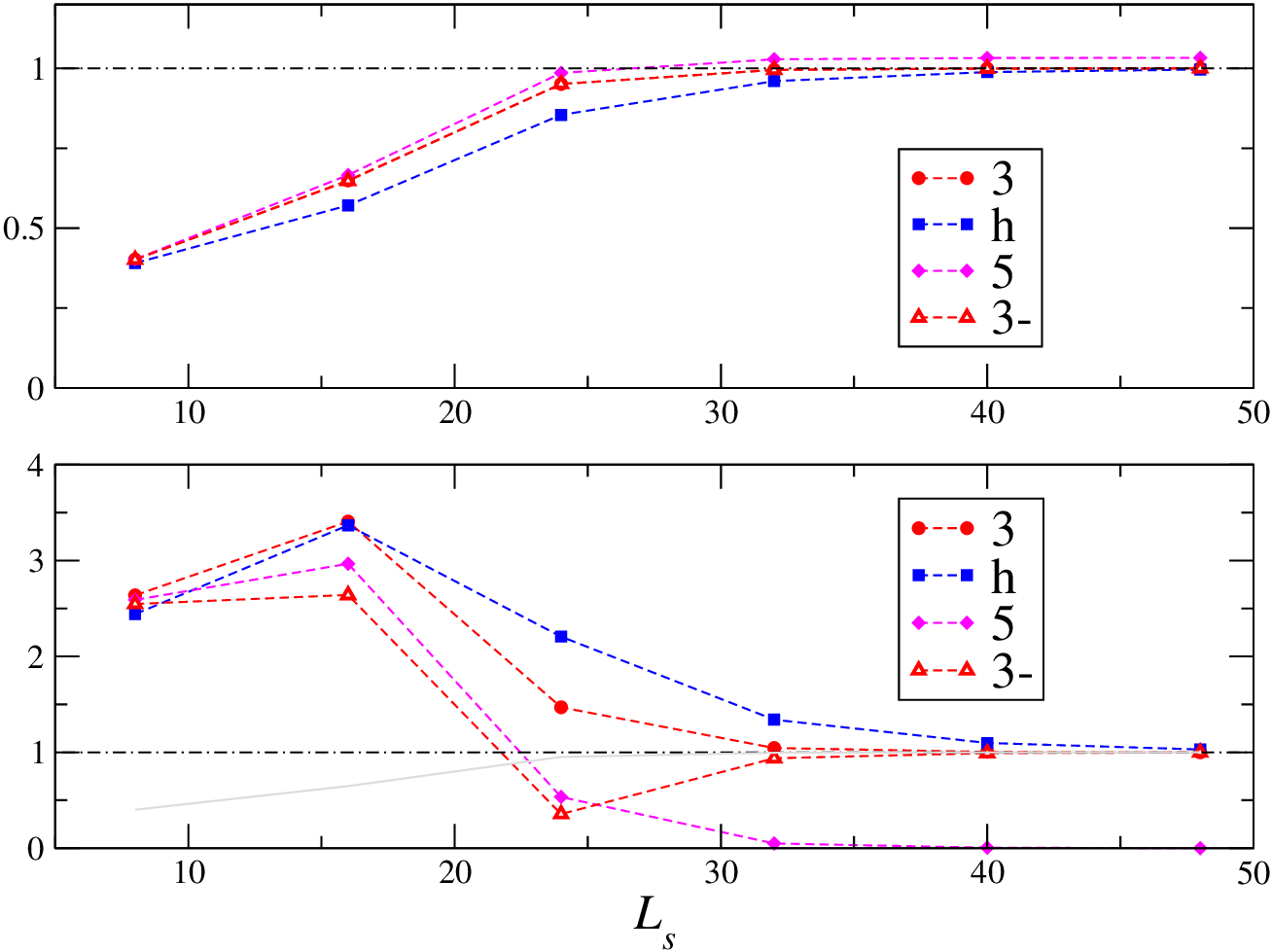}
\caption{Ratio $\sum_{t=0}^{23}\vert C/C^{3\pm-}\vert$ vs. $L_s$ for $C=C^{+-}$ (top)
and $C=C^{--}$ (bottom) for $\beta=0.5$ on $24^3$. The key denotes the mass
term used.}
\label{fig:props_Ls}
\end{center}
\end{figure}
We now present numerical results for the primitive correlators $C^{\pm\pm}(x)$.
It is clear the recovery of U(2) symmetry hinges on the validity of the
approximation (\ref{eq:LargeLsapprox}). As in Sec.~\ref{sec:condensates}, the
first test uses the numerically most demanding  system of $\beta=0.5$ on $24^3\times
L_s$, with fermion mass $m_i=0.01$. For a single equilibrated configuration we calculated $C^{+-}$ and
$C^{--}$, and where relevant their negative mass counterparts $\tilde C^{\pm-}$,
using five randomly chosen sources with $s=0$. Using
$C^{3\pm-}(L_s=48)$ as a reference point, the ratio
$\sum_{t=0}^{23}C(L_s)/C^{3\pm-}(48)$ is
plotted as a function of $L_s$ in Fig.~\ref{fig:props_Ls} for mass terms
$m_3S_3$, $m_hS_h$, $m_5S_5$, along with the same quantity evaluated for $\vert\tilde C^{3}\vert$ (labeled
``3-'' in the plot). It is apparent that $C^{3\pm-}$ 
has reached its large-$L_s$ limit by $L_s\approx40$, but that $C^{h\pm-}$
converges rather more slowly.  Most importantly, $C^{3+-}$ is
indistinguishable from $\tilde C^{3+-}$, whereas $\tilde C^{3--}$ actually changes sign
around $L_s\approx25$ and by $L_s\approx40$ is practically equal to
$-C^{3--}$, consistent with (\ref{eq:LargeLsapprox}). The correlators
$C^{5\pm-}$ do not, however, fit the same pattern; $C^{5+-}$ lies
systematically above $C^{3+-}$, while $C^{5--}\to0$ as $L_s\to\infty$.

\begin{figure}[H]
%\begin{center}
    \centering
    \includegraphics[width=8.5cm]{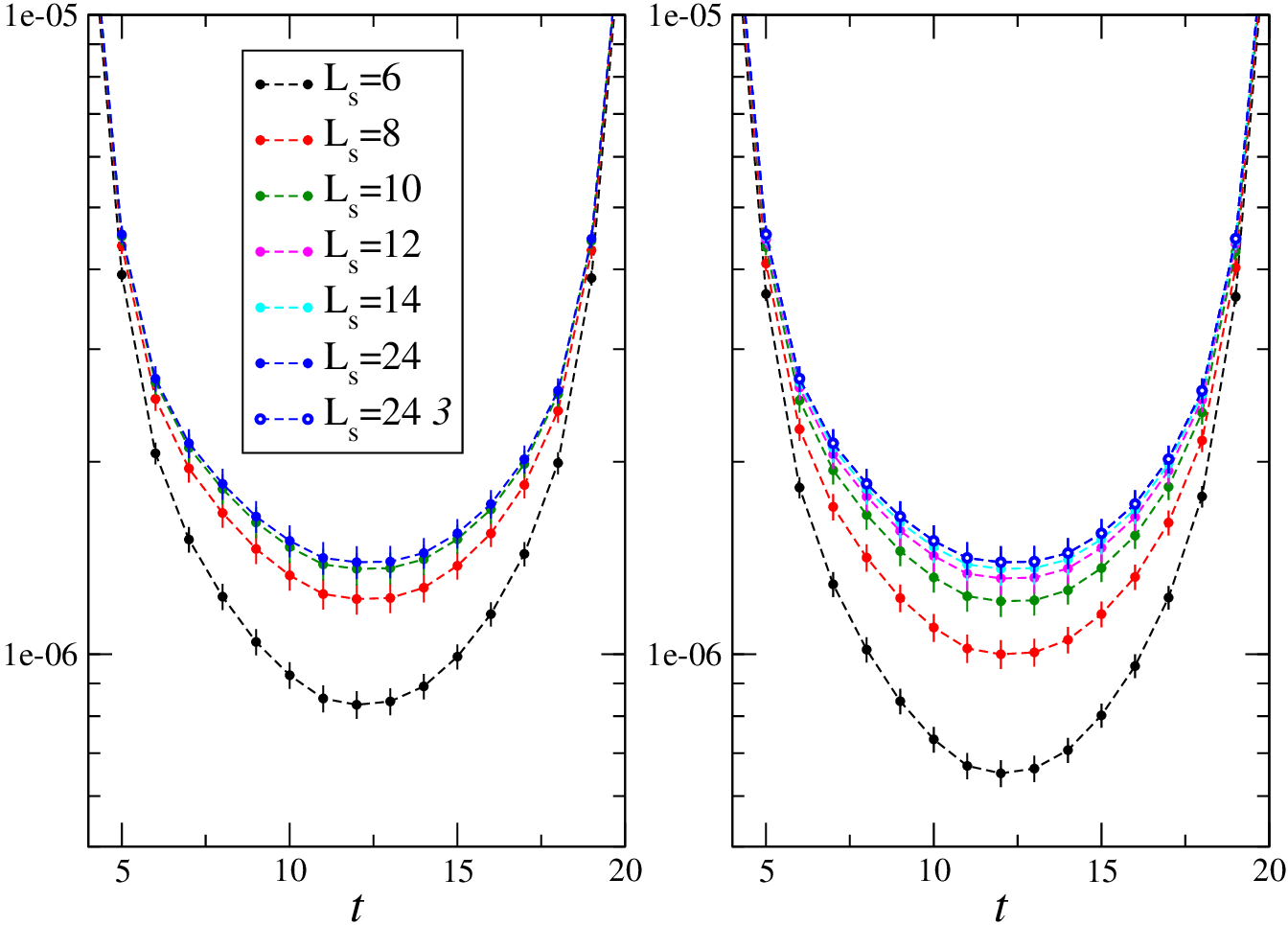}
\caption{Primitive timeslice correlators $C^{3+-}(t)$ (left) and $C^{h+-}(t)$
(right) for various $L_s$ with $\beta=1.0$, $m=0.01$ on $24^3$.}
\label{fig:cpm_Ls}
%\end{center}
\end{figure}
\begin{figure}[thb]
%\begin{center}
    \centering
    \includegraphics[width=8.5cm]{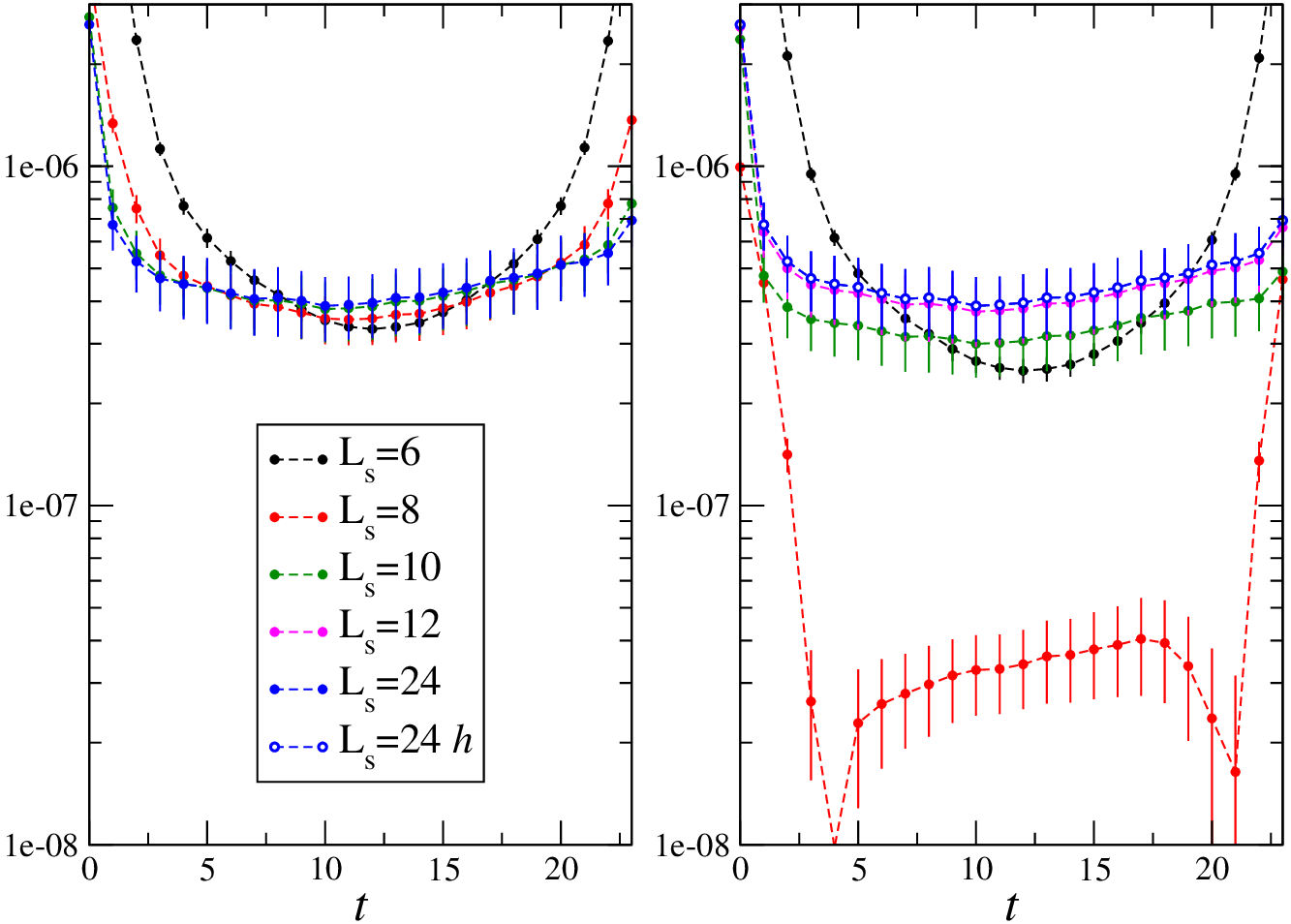}
\caption{Same as Fig.~\ref{fig:cpm_Ls} but for $C^{3--}(t)$ (left) and
$\vert \tilde C^{3--}(t)\vert$ (right).}
\label{fig:cmm_Ls}
%\end{center}
\end{figure}
This picture is confirmed by a study of 500
configurations at $\beta=1.0$, $m_i=0.01$ on $24^3$. Only
the components $C^{\pm-}$,
which are manifestly real, are calculated; the counterparts $C^{\mp+}$ are
identical, and the correlators $\tilde C^{\pm\mp}$ requiring an inversion with a
negative fermion mass recover these properties as $L_s\to\infty$. To focus on
systematic effects the same ensemble was used for each $L_s$ -- the resulting
statistical errorbars show the expected evidence of autocorrelation between
timeslices.
Fig.~\ref{fig:cpm_Ls} compares $C^{3+-}$ with $C^{h+-}$; $C^{3+-}$ converges to
its large-$L_s$ limit for $L_s\gapprox12$, whereas $C^{h+-}$ converges somewhat
more slowly, requiring
$L_s\gapprox16$. Crucially though, by $L_s=24$ the two are indistinguishable.
Fig.~\ref{fig:cmm_Ls} confirms that after a sign change occuring for
$L_s\approx8$ then 
$\lim_{L_s\to\infty}\tilde C^{3--}=-C^{3--}$, and that both also coincide
in magnitude with $C^{h--}$ for $L_s\gapprox24$. We have thus demonstrated that
the mass terms $S_h$ via relations (\ref{eq:Ch3Ch5},\ref{eq:Ch1Ch35}), and $S_3$
via (\ref{eq:C3m3}-\ref{eq:C35m3}), must yield equivalent meson spectra as
$L_s\to\infty$. 

\begin{figure}[tbh]
\begin{center}
    \includegraphics[width=10cm]{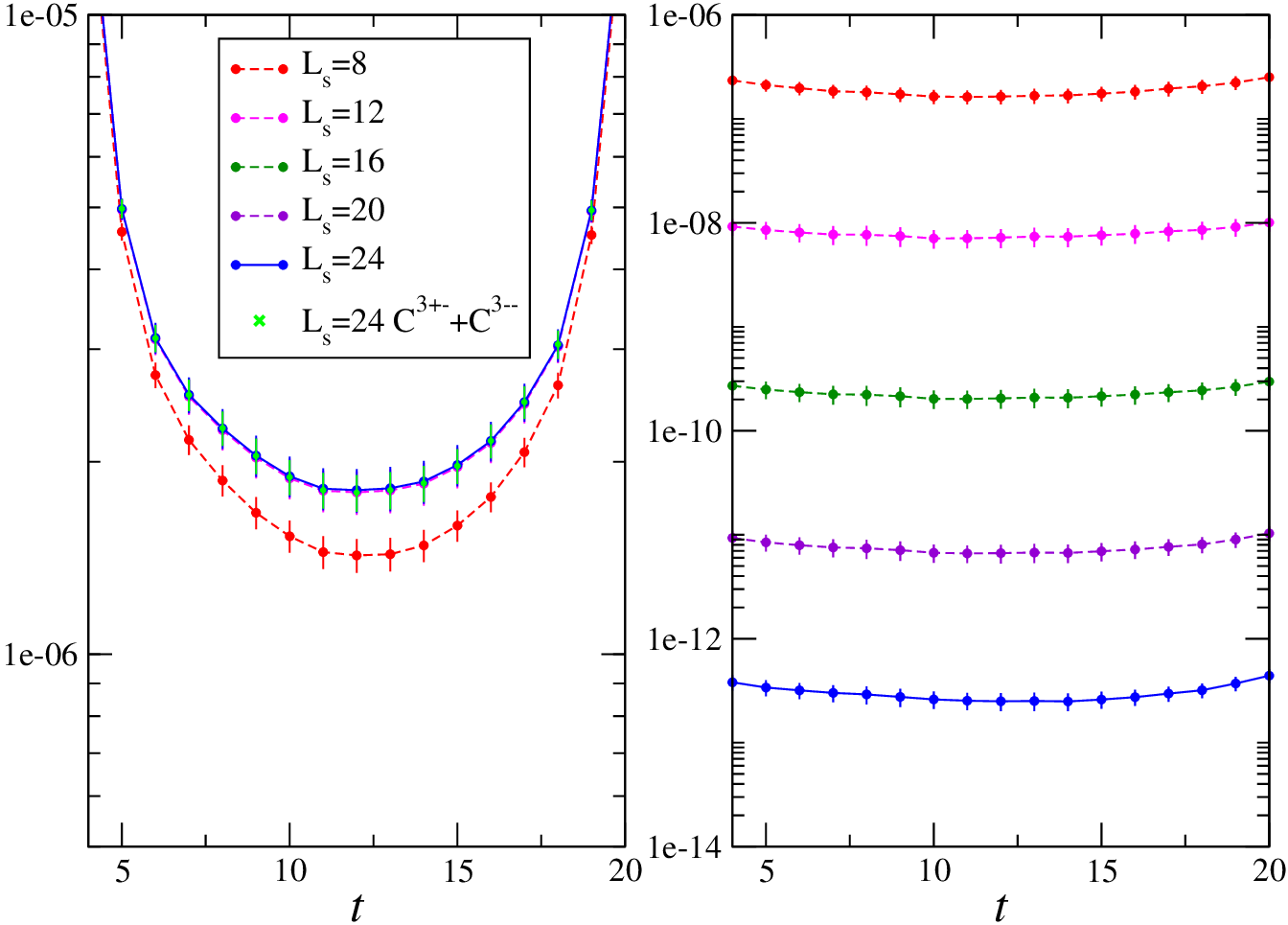}
\caption{Same as Fig.~\ref{fig:cpm_Ls} but for $C^{5+-}(t)$ (left) and
$C^{5--}(t)$ (right). Also shown is the combination $C^{3+-}(t)+C^{3--}(t)$ for
$L_s=24$.}
\label{fig:C5}
\end{center}
\end{figure}
Finally Fig.~\ref{fig:C5} shows the primitive correlators $C^{5\pm-}(t)$ for
various $L_s$. This time the picture is different:
$\lim_{L_s\to\infty}C^{5+-}=C^{3+-}+C^{3--}$, while
$\lim_{L_s\to\infty}C^{5--}=0$. The relations (\ref{eq:C55}-\ref{eq:C355})
then predict that all four meson states become degenerate in this limit, and in this
sense U(2) symmetry is realised. Explicit breaking to U(1)$\otimes$U(1)
by a mass term $m_5S_5$ is not accomplished, however. We can trace this back to the
expressions for the would-be Goldstones interpolated by
$\bar\psi\gamma_3\psi$ (\ref{eq:C35}) and $\bar\psi\psi$ (\ref{eq:C15}) which are
exact even for finite $L_s$, and can only coincide in magnitude in the limit
that either $C^{5--}$ or $C^{5+-}$ vanishes. 
There is thus no opportunity for
degeneracy breaking induced by $S_5$ to develop; 
the system responds to this
constraint by $C^{5+-}$ becoming larger as $L_s\to0$ so that all four mesons become
degenerate with the two would-be Goldstones observed with mass terms $m_hS_h,m_3S_3$. 
Whether this exceptional behaviour is
restricted to a singular point on the U(2) manifold or whether it is a more
general phenomenon must be the subject of further study.

\section{Discussion}
\label{sec:discussion}

The main focus of this paper is the numerical investigation of domain wall
fermions for problems involving fermions in reducible 
spinor representations in 2+1$d$. The results of Sec.~\ref{sec:numbers} suggest
it is highly plausible that the correct global U(2) symmetries are recovered in
the limit $L_s\to\infty$, independently of the continuum limit, just as for
QCD~\cite{Furman:1994ky}. How quickly the
limit is approached remains, of course, a dynamical issue. However, the enhanced
global symmetries of the problem admit new, helpful features not present in QCD, namely 
the ability to introduce a U(2)-breaking mass term in any of three independent
directions all yielding equivalent results as $L_s\to\infty$, aiding
the robust determination of the $L_s$ required in practice. It is also significant
that the dominant artifact $\Delta_h(L_s)$, related to the ``residual mass'' in
QCD simulations, cancels from the physical observables
explored here when the twisted form $im_{3,5}\bar\psi\gamma_{3,5}\psi$ is used.

The only exception to the general good news is the failure of the meson spectrum
calculated with $im_5\bar\psi\gamma_5\psi$ to manifest the expected
U(2)$\to$U(1)$\otimes$U(1) pattern, essentially because the formalism results in 
distinct expressions for the Goldstone correlators in this (possibly
singular) case. The system responds by forcing their difference to vanish as
$L_s\to\infty$, so that no symmetry breakdown occurs. The general conclusion, however, must be that the method looks
extremely promising, and optimal if symmetry breaking is implemented with a mass
term $im_3\bar\psi\gamma_3\psi$.  The next step is to explore a fully
dynamical implementation of domain wall fermions, and then test U(2) restoration by computing
the separate components of the axial Ward identities. In this regard it is worth
noting that the identity (\ref{eq:SvsSdag5}) implies that the fermion
determinant is real for mass terms $S_h$ and $S_3$, and hence there is no
Sign Problem obstruction to running hybrid Monte Carlo
with even $N_f$. For mass term $S_5$ the identity (\ref{eq:SvsSdag3}) does not
suffice to prove $\mbox{det}D^{DW}$ real, but is completely analogous to the
corresponding identity for domain wall fermions in 4$d$. 

An interesting distinction with QCD, or with more general 4$d$ theories, is
the relation between GW/overlap and domain wall approaches.  As 
stressed, the two approaches differ in that GW formulation presented here maintains strict equivalence
between $\gamma_3$ and $\gamma_5$ whereas the domain wall treats them very
differently. We
have seen in Sec.~\ref{sec:GW} that for GW fermions the full U(2) symmetry is only recovered 
in the continuum limit: this is seen either via the effect of the remnant
symmetry rotations $\delta^{GW}$ on bilinears (\ref{eq:GWGold3}), or from the failure of the
fermion propagator identity (\ref{eq:overlapdagger}) to generalise to
$m_{3,5}\not=0$. Perhaps the problem arises from the GW implementation of the
twisted mass terms (\ref{eq:GWtwist}), which as noted introduces an O($a$)
correction which cannot be compensated by a simple field rescaling. On the face
of it, then, GW/overlap fermions can only recover U(2) for $a\to0$, whereas domain wall
fermions apparently recover U(2) for $L_s\to\infty$ irrespective of whether a
continuum limit is taken. One might speculate that there is a different set of GW
relations distinct to (\ref{eq:GW5},\ref{eq:GW3},\ref{eq:GW35}), in which
$\gamma_3$ and $\gamma_5$ do not appear on an equivalent footing, but which is
consistent with the large-$L_s$ limit of domain wall fermions even away from the
continuum limit.
It would also be very valuable to understand things from a
more fundamental perspective by attempting to derive the overlap from the
large-$L_s$ limit of domain wall fermions, as done for 4$d$ in \cite{Neuberger:1997bg}. It will also be
interesting to extend consideration to the case of non-vanishing chemical
potential $\mu$, where there has been some discussion over the correct
implementation of the GW symmetries~\cite{Bloch:2006cd,Gavai:2011np}; happily, there are
several interesting 2+1$d$ systems where simulations at $\mu\not=0$ with orthodox Monte Carlo
methods are feasible~\cite{Hands:2001aq,Hands:2003dh,Armour:2013yk}.

The biggest goal, however, once dynamical fermions have been
implemented, is to study the strongly coupled fixed points of the GN, Thirring
and graphene~\cite{Hands:2008id,Armour:2009vj} systems to see if the critical exponents match those previously
obtained with staggered fermions, and to check whether the expected dependence
on $N_f$ is seen (an exploratory study of the GN model with domain wall fermions
is presented in~\cite{Vranas:1999nx}). This would not only provide an important and hitherto unexplored
test of the applicability of the domain wall approach to lattice fermions, but 
more generally would represent an important step in the definition of
interacting quantum field theories beyond weak coupling. 

\section*{Acknowledgements}
This work was supported by STFC grant ST/L000369/1, and in part by 
National Science Foundation Grant No.
PHYS-1066293 and the hospitality of the Aspen Center for Physics.
The numerical work consumed approximately 7000 Intel Core i5 core hours. 
I have enjoyed discussions with Sam Hawkes, Wolfgang Bietenholz, and in
particular with Tony Kennedy.


\begin{thebibliography}{10}

%\cite{Redlich:1983kn}
\bibitem{Redlich:1983kn}
  A.N.~Redlich,
  %``Gauge Noninvariance and Parity Violation of Three-Dimensional Fermions,''
  Phys.\ Rev.\ Lett.\  {\bf 52} (1984) 18.
  %%CITATION = PRLTA,52,18;%%
  %465 citations counted in INSPIRE as of 27 juil. 2015
%\cite{Redlich:1983dv}
\bibitem{Redlich:1983dv}
  A.N.~Redlich,
  %``Parity Violation and Gauge Noninvariance of the Effective Gauge Field
  %Action in Three-Dimensions,''
  Phys.\ Rev.\ D {\bf 29} (1984) 2366.
  %%CITATION = PHRVA,D29,2366;%%
  %586 citations counted in INSPIRE as of 27 juil. 2015
%\cite{Niemi:1983rq}
\bibitem{Niemi:1983rq}
  A.J.~Niemi and G.W.~Semenoff,
  %``Axial Anomaly Induced Fermion Fractionization and Effective Gauge Theory
  %Actions in Odd Dimensional Space-Times,''
  Phys.\ Rev.\ Lett.\  {\bf 51} (1983) 2077.
  %%CITATION = PRLTA,51,2077;%%
  %427 citations counted in INSPIRE as of 27 juil. 2015

%\cite{Fradkin:1991wy}
\bibitem{Fradkin:1991wy}
  E.~Fradkin and A.~L\'opez,
  %``Fractional Quantum Hall effect and Chern-Simons gauge theories,''
  Phys.\ Rev.\ B {\bf 44} (1991) 5246.
  %%CITATION = PHRVA,B44,5246;%%
  %110 citations counted in INSPIRE as of 27 Jul 2015

%\cite{Pisarski:1984dj}
\bibitem{Pisarski:1984dj}
  R.D.~Pisarski,
  %``Chiral Symmetry Breaking in Three-Dimensional Electrodynamics,''
  Phys.\ Rev.\ D {\bf 29} (1984) 2423.
  %%CITATION = PHRVA,D29,2423;%%
  %292 citations counted in INSPIRE as of 27 Jul 2015

%\cite{Wen:2002zz}
\bibitem{Wen:2002zz}
  X.G.~Wen,
  %``Quantum orders and symmetric spin liquids,''
  Phys.\ Rev.\ B {\bf 65} (2002) 165113.
  %%CITATION = PHRVA,B65,165113;%%
  %98 citations counted in INSPIRE as of 27 Jul 2015
%\cite{Rantner:2002zz}
\bibitem{Rantner:2002zz}
  W.~Rantner and X.G.~Wen,
  %``Spin correlations in the algebraic spin liquid: Implications for high-Tc
  %superconductors,''
  Phys.\ Rev.\ B {\bf 66} (2002) 144501.
  %%CITATION = PHRVA,B66,144501;%%
  %34 citations counted in INSPIRE as of 27 juil. 2015

%\cite{Herbut:2002yq}
\bibitem{Herbut:2002yq}
I.F.~Herbut,
%``QED$_3$ theory of underdoped high temperature superconductors,''
Phys.\ Rev.\ B {\bf 66} (2002) 094504.
%[arXiv:cond-mat/0202491].
%%CITATION = COND-MAT 0202491;%%
%\cite{Franz:2002qy}
\bibitem{Franz:2002qy}
  M.~Franz, Z.~Te\v{s}anovi\'c and O.~Vafek,
%  ``QED$_3$ theory of pairing pseudogap in cuprates. 1. From D wave
%  superconductor to antiferromagnet via 'algebraic' Fermi liquid,''
  Phys.\ Rev.\ B {\bf 66} (2002) 054535.
%  [cond-mat/0203333].
  %%CITATION = COND-MAT/0203333;%%
  %114 citations counted in INSPIRE as of 03 Aug 2015

%\cite{Khveshchenko:2001zz}
\bibitem{Khveshchenko:2001zz}
  D.V.~Khveshchenko,
%  ``Ghost Excitonic Insulator Transition in Layered Graphite,''
  Phys.\ Rev.\ Lett.\  {\bf 87} (2001) 246802.
  %%CITATION = PRLTA,87,246802;%%
  %100 citations counted in INSPIRE as of 03 Aug 2015

%\cite{Son:2007ja}
\bibitem{Son:2007ja}
  D.T.~Son,
%  ``Quantum critical point in graphene approached in the limit of infinitely
%  strong Coulomb interaction,''
  Phys.\ Rev.\ B {\bf75} (2007) 235423.
%  [cond-mat/0701501 [cond-mat.str-el]].
  %%CITATION = COND-MAT/0701501;%%
  %18 citations counted in INSPIRE as of 03 août 2015

\bibitem{CNGPNG}
A.H.~Castro Neto, F.~Guinea, N.M.R.~Peres, K.S.~Novoselov and A.K.~Geim, 
%``The electronic properties of graphene'',
Rev. Mod. Phys. {\bf81} (2009) 109. 
%%CITATION = ARXIV:0709.1163;%%

%\cite{Rosenstein:1990nm}
\bibitem{Rosenstein:1990nm}
  B.~Rosenstein, B.~Warr and S.H.~Park,
%  ``Dynamical symmetry breaking in four Fermi interaction models,''
  Phys.\ Rept.\  {\bf 205} (1991) 59.
  %%CITATION = PRPLC,205,59;%%
  %241 citations counted in INSPIRE as of 27 Jul 2015

%\cite{Gomes:1990ed}
\bibitem{Gomes:1990ed}
  M.~Gomes, R.S.~Mendes, R.F.~Ribeiro and A.J.~da Silva,
%  ``Gauge structure, anomalies and mass generation in a three-dimensional
%  Thirring model,''
  Phys.\ Rev.\ D {\bf 43} (1991) 3516.
  %%CITATION = PHRVA,D43,3516;%%
  %56 citations counted in INSPIRE as of 27 juil. 2015

%\cite{Hands:1992be}
\bibitem{Hands:1992be}
  S.~Hands, A.~Koci\'c and J.B.~Kogut,
%  ``Four Fermi theories in fewer than four-dimensions,''
  Annals Phys.\  {\bf 224} (1993) 29.
%  [hep-lat/9208022].
  %%CITATION = HEP-LAT/9208022;%%
  %121 citations counted in INSPIRE as of 27 juil. 2015

%\cite{Gehring:2015vja}
\bibitem{Gehring:2015vja}
  F.~Gehring, H.~Gies and L.~Janssen,
%  ``Fixed-point structure of low-dimensional relativistic fermion field
%  theories: Universality classes and emergent symmetry,''
  arXiv:1506.07570 [hep-th].
  %%CITATION = ARXIV:1506.07570;%%

\bibitem{G&L}
C.~Gattringer and C.B.~Lang, {\sl Quantum Chromodynamics on the Lattice\/},
(Springer Lecture Notes in Physics {\bf778}, 2010).

%\cite{Coste:1989wf}
\bibitem{Coste:1989wf}
  A.~Coste and M.~L\"uscher,
%  ``Parity Anomaly and Fermion Boson Transmutation in Three-dimensional Lattice
%  {QED},''
  Nucl.\ Phys.\ B {\bf 323} (1989) 631.
  %%CITATION = NUPHA,B323,631;%%
  %61 citations counted in INSPIRE as of 24 juil. 2015

%\cite{Karthik:2015sza}
\bibitem{Karthik:2015sza}
  N.~Karthik and R.~Narayanan,
%  ``Phase of the fermion determinant in QED$_3$ using a gauge invariant lattice
%  regularization,''
  Phys.\ Rev.\ D {\bf 92} (2015) 2,  025003.
%  [arXiv:1505.01051 [hep-th]].
  %%CITATION = ARXIV:1505.01051;%%

%\cite{Burden:1986by}
\bibitem{Burden:1986by}
  C.~Burden and A.N.~Burkitt,
%  ``Lattice Fermions in Odd Dimensions,''
  Europhys.\ Lett.\  {\bf 3} (1987) 545.
  %%CITATION = EULEE,3,545;%%
  %73 citations counted in INSPIRE as of 20 juil. 2015

%\cite{Karkkainen:1993ef}
\bibitem{Karkkainen:1993ef}
  L.~Karkkainen, R.~Lacaze, P.~Lacock and B.~Petersson,
%  ``Critical behavior of the 3-d Gross-Neveu and Higgs-Yukawa models,''
  Nucl.\ Phys.\ B {\bf 415} (1994) 781
   [Nucl.\ Phys.\ B {\bf 438} (1995) 650].
%  [hep-lat/9310020].
  %%CITATION = HEP-LAT/9310020;%%
  %45 citations counted in INSPIRE as of 27 juil. 2015
%\cite{Focht:1995ie}
\bibitem{Focht:1995ie}
  E.~Focht, J.~Jers\'ak and J.~Paul,
%  ``Interplay of universality classes in a three-dimensional Yukawa model,''
  Phys.\ Rev.\ D {\bf 53} (1996) 4616.
%  [hep-lat/9511005].
  %%CITATION = HEP-LAT/9511005;%%
  %19 citations counted in INSPIRE as of 27 Jul 2015

%\cite{DelDebbio:1997dv}
\bibitem{DelDebbio:1997dv}
  L.~Del Debbio, S.~Hands and J.C.~Mehegan,
%  ``The Three-dimensional Thirring model for small $N_f$,''
  Nucl.\ Phys.\ B {\bf 502} (1997) 269.
%  [hep-lat/9701016].
  %%CITATION = HEP-LAT/9701016;%%
  %48 citations counted in INSPIRE as of 27 Jul 2015
%\cite{Frick:1994ry}
\bibitem{Frick:1994ry}
  C.~Frick and J.~Jers\'ak,
%  ``Dynamical fermion mass generation by strong gauge interaction shielded by a
%  scalar field,''
  Phys.\ Rev.\ D {\bf 52} (1995) 340.
%  [hep-lat/9411084].
  %%CITATION = HEP-LAT/9411084;%%
  %23 citations counted in INSPIRE as of 27 Jul 2015

%\cite{Christofi:2007ye}
\bibitem{Christofi:2007ye}
  S.~Christofi, S.~Hands and C.~Strouthos,
%  ``Critical flavor number in the three dimensional Thirring model,''
  Phys.\ Rev.\ D {\bf 75} (2007) 101701.
%  [hep-lat/0701016].
  %%CITATION = HEP-LAT/0701016;%%
  %32 citations counted in INSPIRE as of 27 Jul 2015

%\cite{Chandrasekharan:2011mn}
\bibitem{Chandrasekharan:2011mn}
  S.~Chandrasekharan and A.~Li,
%  ``Fermion bags, duality and the three dimensional massless lattice Thirring
%  model,''
  Phys.\ Rev.\ Lett.\  {\bf 108} (2012) 140404.
%  [arXiv:1111.7204 [hep-lat]].
  %%CITATION = ARXIV:1111.7204;%%
  %30 citations counted in INSPIRE as of 27 juil. 2015
%\cite{Chandrasekharan:2013aya}
\bibitem{Chandrasekharan:2013aya}
  S.~Chandrasekharan and A.~Li,
%  ``Quantum critical behavior in three dimensional lattice Gross-Neveu
%  models,''
  Phys.\ Rev.\ D {\bf 88} (2013) 021701.
%  [arXiv:1304.7761 [hep-lat]].
  %%CITATION = ARXIV:1304.7761;%%
  %12 citations counted in INSPIRE as of 27 juil. 2015

%\cite{Appelquist:1999hr}
\bibitem{Appelquist:1999hr}
  T.~Appelquist, A.G.~Cohen and M.~Schmaltz,
%  ``A New constraint on strongly coupled gauge theories,''
  Phys.\ Rev.\ D {\bf 60} (1999) 045003.
%  [hep-th/9901109].
  %%CITATION = HEP-TH/9901109;%%
  %138 citations counted in INSPIRE as of 27 Jul 2015

%\cite{Aleiner:2007va}
\bibitem{Aleiner:2007va}
  I.L.~Aleiner, D.E.~Kharzeev and A.M.~Tsvelik,
%  ``Spontaneous symmetry breakings in graphene subjected to in-plane magnetic
%  field,''
  Phys.\ Rev.\ B {\bf 76} (2007) 195415.
%  [arXiv:0708.0394 [cond-mat.mes-hall]].
  %%CITATION = ARXIV:0708.0394;%%
  %24 citations counted in INSPIRE as of 27 juil. 2015

%\cite{Ginsparg:1981bj}
\bibitem{Ginsparg:1981bj}
  P.H.~Ginsparg and K.G.~Wilson,
%  ``A Remnant of Chiral Symmetry on the Lattice,''
  Phys.\ Rev.\ D {\bf 25} (1982) 2649.
  %%CITATION = PHRVA,D25,2649;%%
  %909 citations counted in INSPIRE as of 20 juil. 2015

%\cite{Kaplan:1992bt}
\bibitem{Kaplan:1992bt}
  D.B.~Kaplan,
%  ``A Method for simulating chiral fermions on the lattice,''
  Phys.\ Lett.\ B {\bf 288} (1992) 342.
%  [hep-lat/9206013].
  %%CITATION = HEP-LAT/9206013;%%
  %957 citations counted in INSPIRE as of 20 juil. 2015

%\cite{Furman:1994ky}
\bibitem{Furman:1994ky}
  V.~Furman and Y.~Shamir,
%  ``Axial symmetries in lattice QCD with Kaplan fermions,''
  Nucl.\ Phys.\ B {\bf 439} (1995) 54.
%  [hep-lat/9405004].
  %%CITATION = HEP-LAT/9405004;%%
  %505 citations counted in INSPIRE as of 20 juil. 2015

%\cite{Neuberger:1997fp}
\bibitem{Neuberger:1997fp}
  H.~Neuberger,
%  ``Exactly massless quarks on the lattice,''
  Phys.\ Lett.\ B {\bf 417} (1998) 141.
%  [hep-lat/9707022].
  %%CITATION = HEP-LAT/9707022;%%
%\cite{Neuberger:1998wv}
\bibitem{Neuberger:1998wv}
  H.~Neuberger,
%  ``More about exactly massless quarks on the lattice,''
  Phys.\ Lett.\ B {\bf 427} (1998) 353.
%  [hep-lat/9801031].
  %%CITATION = HEP-LAT/9801031;%%
  %669 citations counted in INSPIRE as of 24 juil. 2015

\bibitem{Hands:1989mv}
  S.~Hands and J.B.~Kogut,
%  ``Finite Size Effects and Chiral Symmetry Breaking in Quenched
%  Three-dimensional {QED},''
  Nucl.\ Phys.\ B {\bf 335} (1990) 455.
  %%CITATION = NUPHA,B335,455;%%
  %37 citations counted in INSPIRE as of 03 Jul 2015mabegin{equation}

%\cite{Frezzotti:2000nk}
\bibitem{Frezzotti:2000nk}
  R.~Frezzotti, P.~Grassi, S.~Sint and P.~Weisz,
%  ``Lattice QCD with a chirally twisted mass term,''
  JHEP {\bf 0108} (2001) 058.
  %%CITATION = HEP-LAT/0101001;%%
  %356 citations counted in INSPIRE as of 24 juil. 2015

%\cite{Herbut:2009qb}
\bibitem{Herbut:2009qb}
  I.F.~Herbut, V.~Juri\v ci\'c and B.~Roy,
%  ``Theory of interacting electrons on the honeycomb lattice,''
  Phys.\ Rev.\ B {\bf 79} (2009) 085116.
%  [arXiv:0811.0610 [cond-mat.str-el]].
  %%CITATION = ARXIV:0811.0610;%%
  %65 citations counted in INSPIRE as of 23 juil. 2015

%\cite{Semenoff:1984dq}
\bibitem{Semenoff:1984dq}
  G.W.~Semenoff,
%  ``Condensed Matter Simulation of a Three-dimensional Anomaly,''
  Phys.\ Rev.\ Lett.\  {\bf 53} (1984) 2449.
  %%CITATION = PRLTA,53,2449;%%
  %296 citations counted in INSPIRE as of 03 Aug 2015

%\cite{Hou:2006qc}
\bibitem{Hou:2006qc}
  C.Y.~Hou, C.~Chamon and C.~Mudry,
%  ``Electron fractionalization in two-dimensional graphene-like structures,''
  Phys.\ Rev.\ Lett.\  {\bf 98} (2007) 186809.
%  [cond-mat/0609740 [cond-mat.mes-hall]].
  %%CITATION = COND-MAT/0609740;%%
  %81 citations counted in INSPIRE as of 03 août 2015

%\cite{Haldane:1988zza}
\bibitem{Haldane:1988zza}
  F.D.M.~Haldane,
%  ``Model for a Quantum Hall Effect without Landau Levels: Condensed-Matter
%  Realization of the 'Parity Anomaly',''
  Phys.\ Rev.\ Lett.\  {\bf 61} (1988) 2015.
  %%CITATION = PRLTA,61,2015;%%
  %330 citations counted in INSPIRE as of 03 Aug 2015

%\cite{Nielsen:1980rz}
\bibitem{Nielsen:1980rz}
  H.B.~Nielsen and M.~Ninomiya,
%  ``Absence of Neutrinos on a Lattice. 1. Proof by Homotopy Theory,''
  Nucl.\ Phys.\ B {\bf 185} (1981) 20
   [Nucl.\ Phys.\ B {\bf 195} (1982) 541].
  %%CITATION = NUPHA,B185,20;%%
  %910 citations counted in INSPIRE as of 20 juil. 2015
%\cite{Nielsen:1981xu}
\bibitem{Nielsen:1981xu}
  H.B.~Nielsen and M.~Ninomiya,
%  ``Absence of Neutrinos on a Lattice. 2. Intuitive Topological Proof,''
  Nucl.\ Phys.\ B {\bf 193} (1981) 173.
  %%CITATION = NUPHA,B193,173;%%
  %585 citations counted in INSPIRE as of 20 juil. 2015

%\cite{Bietenholz:2000ca}
\bibitem{Bietenholz:2000ca}
  W.~Bietenholz and J.~Nishimura,
%  ``Ginsparg-Wilson fermions in odd dimensions,''
  JHEP {\bf 0107} (2001) 015.
%  [hep-lat/0012020].
  %%CITATION = HEP-LAT/0012020;%%
  %16 citations counted in INSPIRE as of 24 Jul 2015

%\cite{Kikukawa:1997qh}
\bibitem{Kikukawa:1997qh}
  Y.~Kikukawa and H.~Neuberger,
%  ``Overlap in odd dimensions,''
  Nucl.\ Phys.\ B {\bf 513} (1998) 735.
%  [hep-lat/9707016].
  %%CITATION = HEP-LAT/9707016;%%
  %51 citations counted in INSPIRE as of 24 juil. 2015

%\cite{Luscher:1998pqa}
\bibitem{Luscher:1998pqa}
  M.~L\"uscher,
%  ``Exact chiral symmetry on the lattice and the Ginsparg-Wilson relation,''
  Phys.\ Lett.\ B {\bf 428} (1998) 342.
%  [hep-lat/9802011]
  %%CITATION = HEP-LAT/9802011;%%
  %710 citations counted in INSPIRE as of 28 juil. 2015

%\cite{Hands:2004bh}
\bibitem{Hands:2004bh}
  S.J.~Hands, J.B.~Kogut, L.~Scorzato and C.G.~Strouthos,
%  ``Non-compact QED(3) with $N_f=1$ and $N_f=4$,''
  Phys.\ Rev.\ B {\bf 70} (2004) 104501. 
%  [hep-lat/0404013]
  %%CITATION = HEP-LAT/0404013;%%
  %59 citations counted in INSPIRE as of 13 Jul 2015

%\cite{Neuberger:1997bg}
\bibitem{Neuberger:1997bg}
  H.~Neuberger,
%  ``Vector - like gauge theories with almost massless fermions on the
%  lattice,''
  Phys.\ Rev.\ D {\bf 57} (1998) 5417.
%  [hep-lat/9710089].
  %%CITATION = HEP-LAT/9710089;%%
  %290 citations counted in INSPIRE as of 27 juil. 2015

%\cite{Bloch:2006cd}
\bibitem{Bloch:2006cd}
  J.C.R.~Bloch and T.~Wettig,
%  ``Overlap Dirac operator at nonzero chemical potential and random matrix
%  theory,''
  Phys.\ Rev.\ Lett.\  {\bf 97} (2006) 012003.
%  [hep-lat/0604020].
  %%CITATION = HEP-LAT/0604020;%%
  %63 citations counted in INSPIRE as of 27 juil. 2015
%\cite{Gavai:2011np}
\bibitem{Gavai:2011np}
  R.V.~Gavai and S.~Sharma,
%  ``Exact chiral invariance at finite density on the lattice,''
  Phys.\ Lett.\ B {\bf 716} (2012) 446.
%  [arXiv:1111.5944 [hep-lat]].
  %%CITATION = ARXIV:1111.5944;%%
  %8 citations counted in INSPIRE as of 27 juil. 

%\cite{Hands:2001aq}
\bibitem{Hands:2001aq}
  S.~Hands, B.~Lucini and S.~Morrison,
%  ``Numerical portrait of a relativistic thin film BCS superfluid,''
  Phys.\ Rev.\ D {\bf 65} (2002) 036004.
%  [hep-lat/0109001].
  %%CITATION = HEP-LAT/0109001;%%
  %32 citations counted in INSPIRE as of 27 juil. 2015
%\cite{Hands:2003dh}
\bibitem{Hands:2003dh}
  S.~Hands, J.B.~Kogut, C.G.~Strouthos and T.N.~Tran,
%  ``Fermi surface phenomena in the (2+1)-d four Fermi model,''
  Phys.\ Rev.\ D {\bf 68} (2003) 016005.
%  [hep-lat/0302021].
  %%CITATION = HEP-LAT/0302021;%%
  %21 citations counted in INSPIRE as of 27 juil. 2015
%\cite{Armour:2013yk}
\bibitem{Armour:2013yk}
  W.~Armour, S.~Hands and C.~Strouthos,
%  ``Monte Carlo study of strongly interacting degenerate fermions: A model for
%  voltage-biased bilayer graphene,''
  Phys.\ Rev.\ D {\bf 87} (2013),  065010.
%  [arXiv:1302.0150 [hep-lat]].
  %%CITATION = ARXIV:1302.0150;%%

%\cite{Hands:2008id}
\bibitem{Hands:2008id}
  S.~Hands and C.~Strouthos,
%  ``Quantum Critical Behaviour in a Graphene-like Model,''
  Phys.\ Rev.\ B {\bf 78} (2008) 165423.
%  [arXiv:0806.4877 [cond-mat.str-el]].
  %%CITATION = ARXIV:0806.4877;%%
  %66 citations counted in INSPIRE as of 27 Jul 2015
%\cite{Armour:2009vj}
\bibitem{Armour:2009vj}
  W.~Armour, S.~Hands and C.~Strouthos,
%  ``Monte Carlo Simulation of the Semimetal-Insulator Phase Transition in
%  Monolayer Graphene,''
  Phys.\ Rev.\ B {\bf 81} (2010) 125105.
%  [arXiv:0910.5646 [cond-mat.str-el]].
  %%CITATION = ARXIV:0910.5646;%%
  %66 citations counted in INSPIRE as of 27 juil. 2015

%\cite{Vranas:1999nx}
\bibitem{Vranas:1999nx}
  P.~Vranas, I.~Tziligakis and J.B.~Kogut,
%  ``Fermion scalar interactions with domain wall fermions,''
  Phys.\ Rev.\ D {\bf 62} (2000) 054507.
%  [hep-lat/9905018].
  %%CITATION = HEP-LAT/9905018;%%
  %11 citations counted in INSPIRE as of 03 août 2015

\end{thebibliography}
\end{document}